\documentclass{article}

\usepackage[nonatbib, preprint]{neurips_data_2024}

\usepackage[square,numbers]{natbib}
\usepackage{fontawesome}
\usepackage{hyperref}
\usepackage{url}
\usepackage{wrapfig}
\usepackage{subfigure}
\usepackage{xcolor}
\usepackage{booktabs}
\usepackage{adjustbox}
\usepackage{graphicx}
\usepackage{bm} 
\usepackage{float}
\usepackage[inkscapearea=page]{svg}
\usepackage{tocbibind}
\usepackage[toc,page]{appendix} 

\newenvironment{tightemize}
{\begin{itemize}\itemsep1pt \parskip0pt \parsep0pt}{\end{itemize}\vspace{-\topsep}}
\usepackage[utf8]{inputenc} 
\usepackage[T1]{fontenc}    
\usepackage{hyperref}       
\usepackage{url}            
\usepackage{booktabs}       
\usepackage{amsfonts}       
\usepackage{nicefrac}       
\usepackage{microtype}      
\usepackage{xcolor}         
\usepackage{amsmath} 
\usepackage{titlesec}

\titlespacing\section{0pt}{1pt plus 6pt minus 2pt}{1pt plus 2pt minus 1pt}
\titlespacing\subsection{0pt}{8pt plus 3pt minus 2pt}{4pt plus 1pt minus 1pt}
\titlespacing\subsubsection{0pt}{6pt plus 1pt minus 1pt}{3pt plus 1pt minus 1pt}

\expandafter\def\expandafter\normalsize\expandafter{%
    \normalsize%
    \setlength\abovedisplayskip{0pt}%
    \setlength\belowdisplayskip{8pt}%
    \setlength\abovedisplayshortskip{-8pt}%
    \setlength\belowdisplayshortskip{2pt}%
}

\title{Antibody DomainBed: Out-of-Distribution Generalization in Therapeutic Protein Design}

\vspace{-0.5cm}
\author{Nata\v{s}a Tagasovska\thanks{equal contribution }\thanks{Prescient Design, Genentech} 
\And 
Ji Won Park \footnotemark[1] \footnotemark[2] \\
\And
Matthieu Kirchmeyer \footnotemark[2]
\And Nathan C. Frey \footnotemark[2]
\And Andrew Martin Watkins \footnotemark[2]
\And Aya Abdelsalam Ismail \footnotemark[2]
\And
Arian Rokkum Jamasb\footnotemark[2]
\And Edith Lee \footnotemark[2]
\And
Tyler Bryson\footnotemark[2]
\And 
Stephen Ra \footnotemark[2]
\And
Kyunghyun Cho\footnotemark[2]\thanks{Department of Computer Science, Center for Data Science, New York University} \\
}

\begin{document}

\maketitle
\begin{abstract}
    Machine learning (ML) has demonstrated significant promise in accelerating drug design. Active ML-guided optimization of therapeutic molecules typically relies on a surrogate model predicting the target property of interest. The model predictions are used to determine which designs to evaluate in the lab, and the model is updated on the new measurements to inform the next cycle of decisions. A key challenge is that the experimental feedback from each cycle inspires changes in the candidate proposal or experimental protocol for the next cycle, which lead to distribution shifts. To promote robustness to these shifts, we must account for them explicitly in the model training. We apply domain generalization (DG) methods to classify the stability of interactions between an antibody and antigen across five domains defined by design cycles. Our results suggest that foundational models and ensembling improve predictive performance on out-of-distribution domains. We publicly release our codebase extending the DG benchmark ``DomainBed,'' and the associated dataset of antibody sequences and structures emulating distribution shifts across design cycles. 
  
\end{abstract}
   
\section{Introduction}
\label{sec:intro}

A model trained to minimize training error is incentivized to capture all correlations in the training data. If the training and test data are sampled from different distributions, however, this can lead to catastrophic failures outside the training domain \citep{torralba2011unbiased, zech2018variable, beery2019efficient, koh2021wilds, neuhaus2022spurious}. Domain generalization (DG) aims to build robust predictors that generalize to unseen test domains. A common DG approach extracts domain invariance from datasets spanning multiple training domains \citep{blanchard2011generalizing, muandet2013domain, arjovsky2019invariant}. Inspired by causality, this work seeks to isolate causal factors of variation, which are stable across domains, from spurious ones that may change between training and test domains \citep{arjovsky2019invariant, ahuja2021invariance, rame2022fishr}.

Benchmarking efforts for DG algorithms, to date, have been largely limited to image classification tasks \citep[e.g.,][]{gulrajani2020search, lynch2023spawrious}. To prepare these algorithms for critical applications such as healthcare and medicine, we must validate and stress-test them on a wide variety of real-world datasets carrying selection biases, confounding factors, and other domain-specific idiosyncrasies. To our knowledge, this paper is the first to apply DG algorithms to the problem of active drug design, a setting riddled with complex distribution shifts. 

\begin{wrapfigure}{r}{0.35\textwidth}
    \begin{center}
    \includegraphics[width=0.35\textwidth]{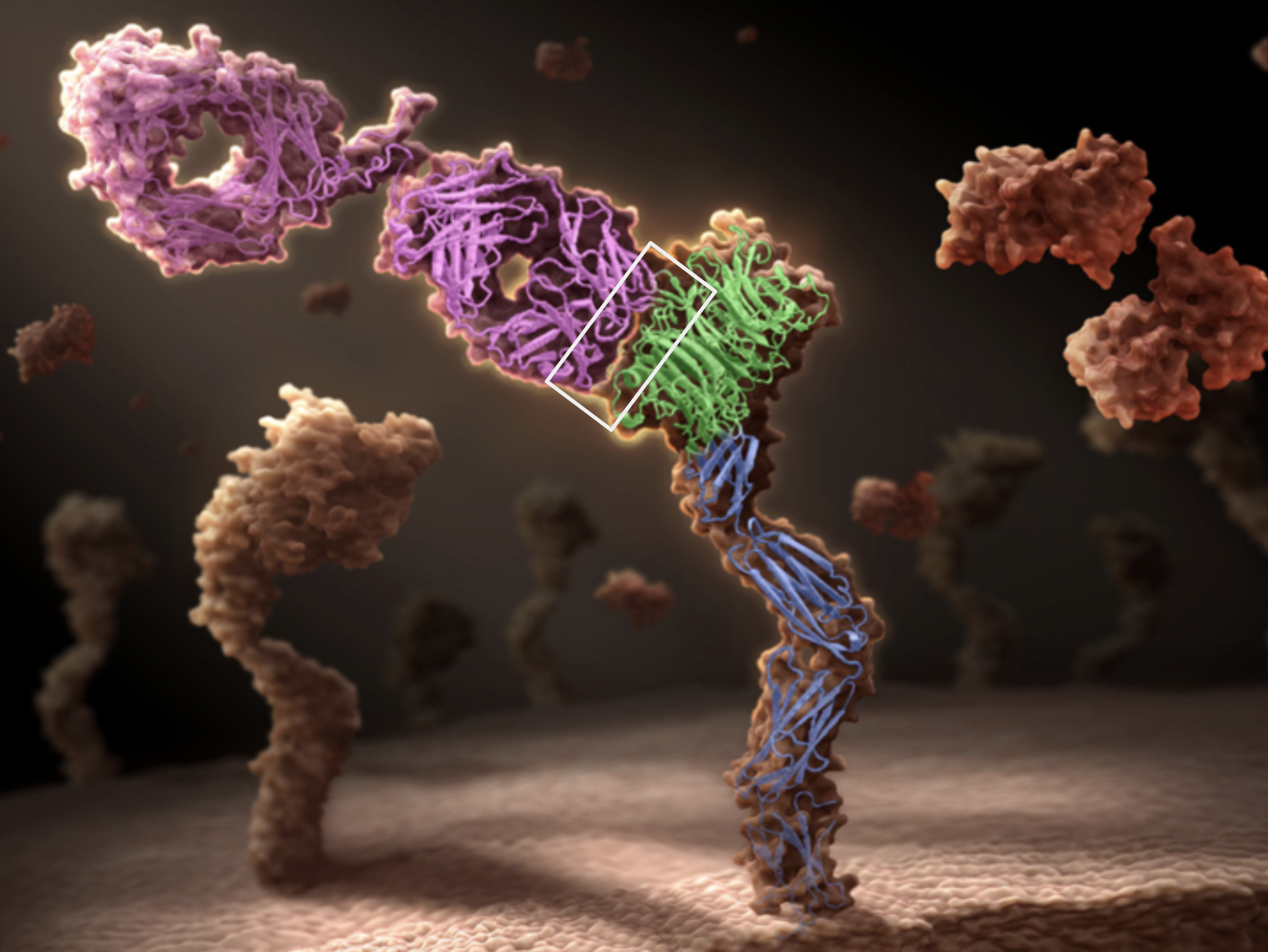}
        \caption{\label{fig:context} \textbf{Stabilizing interactions between antibody and antigen.} Antibody Onartuzumab\protect\footnote{in late-stage clinical trials targeting multiple types of lung cancer, \citep{rolfo2015onartuzumab}} (pink) binds to MET (green and blue), a lung cancer antigen, on the cell surface. The strength of binding is determined by the binding site of the antibody interacting with the antigen, boxed in white.}
    \end{center}
    \vspace{-1cm}
\end{wrapfigure}

The specific application we consider is that of characterizing the interactions between therapeutic antibodies and target antigens. Antibodies are proteins used by the immune system to recognize harmful substances (antigens) such as bacteria and viruses \citep{singh2018monoclonal}. They bind, or attach, to antigens in order to mediate an immune response against them. The strength of binding is determined by the binding site of the antibody (paratope) interacting with the antigen epitope (\autoref{fig:context}). Antibodies that bind tightly to a given target antigen are highly desirable as therapeutic candidates. 

The lab experiments that measure the binding affinity of antibodies are costly and time-consuming. In active antibody design, we thus assign a surrogate ML model to predict binding and select the most promising candidates for wet-lab evaluation based on the predictions. Developing an accurate surrogate model is a challenging task in itself, because, as explained in more detail in \autoref{sec:mldd}, the model may latch onto \textbf{non-mechanistic} factors of variation in the data that do not cause binding: identity of the target antigen, assay used to measure binding, distinguishing features of the generative models (either human experts or ML) that proposed the antibody, and ``batch effects'' that create heteroscedastic measurement errors.

\textbf{Contributions and Impact.} We bring a novel perspective to active drug design by applying domain generalization (DG) methodologies. The iterative nature of active drug design, characterized by significant distribution shifts, provides a rich and previously underexplored testbed for DG algorithms. In turn, the field of drug design stands to gain immensely from robust surrogate binding predictors developed through DG techniques. This synergy enables two key outcomes: (1) providing impactful, real-world benchmarking for DG algorithms and (2) developing robust predictors that can significantly enhance active antibody design.
\begin{tightemize}
    \item We open-source a joint sequence and structure dataset for OOD generalization for antibodies emulating active drug design.
    \item We review and evaluate the latest DG algorithms in the context of active drug design, the first large-scale benchmark study on large molecules to the best of our knowledge.
    \item We present best-practice guidelines and highlight open questions for the field. 
\end{tightemize}


\section{Accelerating antibody design with ML}
\label{sec:mldd}
\textbf{Problem formulation.}
Antibody design typically focuses on designing the variable region of an antibody, which consists of the heavy chain and the light chain. Each chain can be represented as a sequence of characters from an alphabet of $20$ characters (for $20$ possible amino acids). The heavy and light chains combined span $L \sim 290$ amino acids on average. 
We denote the sequences as ${\bm x} = (a_1, \ldots, a_L)$, where $a_l \in \{1, \ldots, 20\}$ 
corresponds to the amino acid type at position $l \in [L]$. 
We experimentally measure the binding affinity $z \in \mathbb{R}$ from each sequence. For simplicity, we create a classification task by creating a binary label $y \in \{0, 1\}$ from $z$. We set $y=1$ if $z$ exceeds a chosen minimum affinity value that would qualify as binding and $y=0$ otherwise. Each antibody ${\bm x}_i$, indexed $i$, carries a label $y_i$ in one of the $N_r=5$ design rounds - $r$, where $r \in \lbrace 0, \dots , 4 \rbrace$.
The labeled dataset for a round $r$ is a set of $n_r$ ordered pairs: $\mathcal{D}_r = \{({\bm {x}}_i^r, y_i^r)\}_{i=1}^{n_r}$. 

\begin{figure}[t!]
    \centering
    \adjustbox{trim={3.5cm 1cm 2cm 0}}{
    \includegraphics[width=\textwidth]{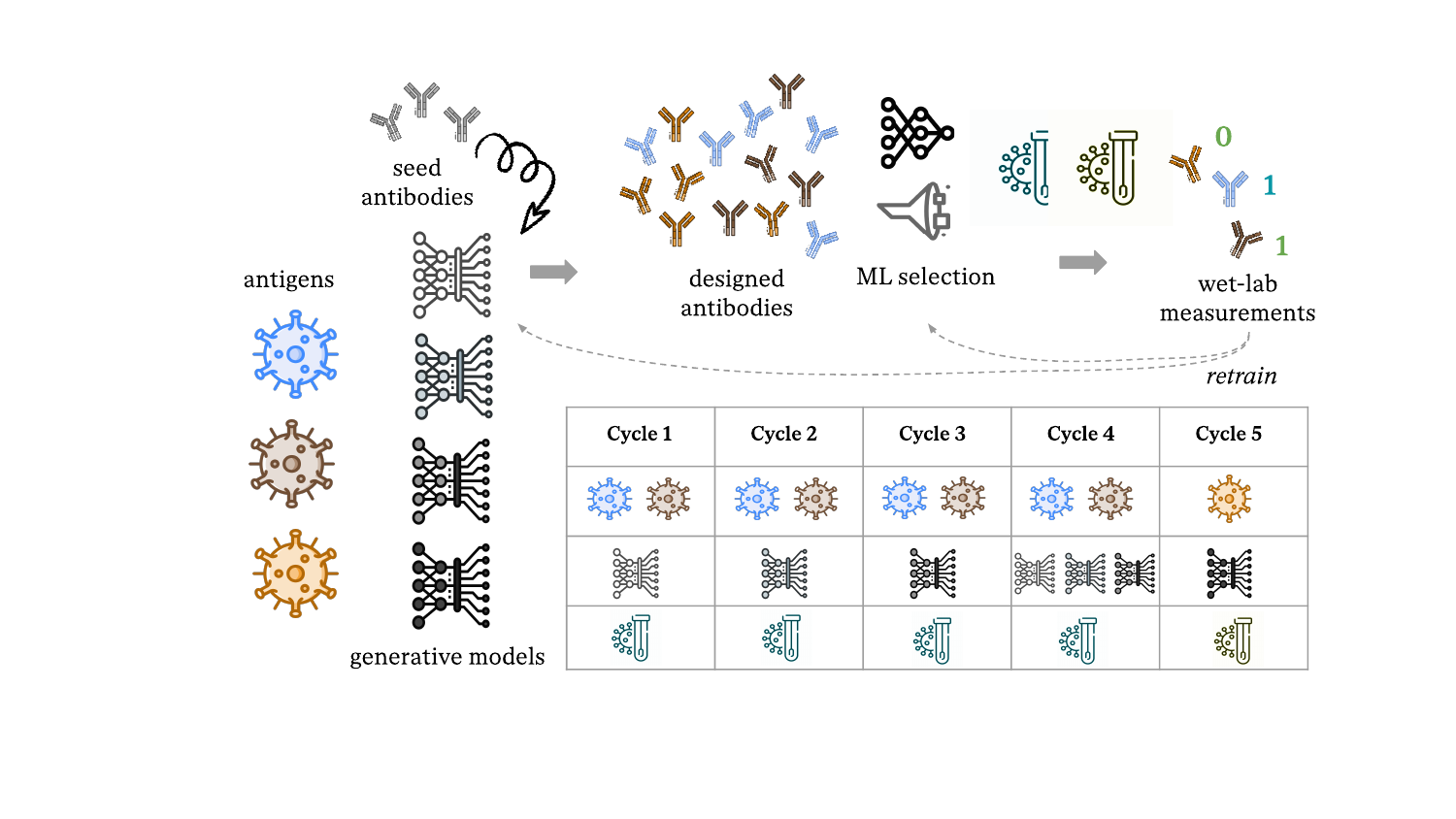}}
    \vspace{-0.5cm}
     \caption{Active ML-guided design of antibodies against specific antigens involves: (1) developing generative models to create novel antibodies from seed antibodies, (2) selecting promising designs with predictive models, (3) experimentally validating the designs, and (4) updating models with new data for the next cycle. Each cycle may vary in targets, generative models, and experimental assays.}
    \label{fig:intro}
     \vspace{-0.5cm}
 \end{figure}

\textbf{Lab in the loop.}
\label{sec:lll}
Our antibody binding dataset is generated from an active ML-guided design process involving multiple design cycles, or \emph{rounds}. As illustrated in \autoref{fig:intro}, each round consists of the following steps:
\textbf{1}. Millions of candidate sequences are sampled from a suite of generative models, including variational autoencoders \citep{gligorijevic2021function, berenberg2022multi}, energy-based models \citep{tagasovska2022pareto, frey2023learning} and diffusion models \citep{gruver2023protein, frey2023protein}. 
\textbf{2}. A small subset of several hundred promising candidates is selected based on binding predictions from a surrogate binding classifier \citep{park2022propertydag}.
\textbf{3}. The lab experimentally measures binding.
\textbf{4}. All models (generative and discriminative) are updated with new measurements.
In Step 4, both the generative model and the surrogate classifier $\hat{f}_{\theta}$ are updated. Beyond being refit on the new data returned from the lab, the generative models may undergo more fundamental modifications in their architectures, pretrained weights, and training schemes. 

\textbf{Challenge.} A standard approach to supervised learning tasks is empirical risk minimization (ERM) \citep{vapnik1992principles}. Let us first define the risk in each round $r$ as
\begin{align} \label{eq:per_Env risk}
    \mathcal{R}^r(\theta) = \mathbb{E}_{(X^r, Y^r)\sim \mathcal{D}_{r_j} }  \ell \left(\hat{f}_\theta(X^r), Y^r \right),
\end{align}
where $\ell$ is the loss function. ERM simply minimizes the training error, i.e., the average risk across all the training examples from all the rounds.
\begin{align} \label{eq:erm_risk}
  \mathcal{R}_{\rm ERM}(\theta) &= \mathbb{E}_{(X^r, Y^r)\sim { \bigcup_{j\in N_r }} \mathcal{D}_{r_j} }  \ell \left(\hat{f}_\theta(X^r), Y^r \right)  = \mathbb{E}_{r \sim {p_{\rm train}(r)}} \mathcal{R}^r(\theta),
\end{align}
where $p_{\rm train}(r)$ denotes distribution of the rounds in the training set. When we trained our surrogate classifier by ERM, it did not improve significantly even as the training set size increased over design rounds. In each subsequent round, representing the test domain, we observed that the performance approaches random, indicating large distribution shifts. In particular, in analysis in \autoref{sec:results} we reproduce such result, and we show that classifiers, regardless of the architecture and data size involved, indeed latch to spurious correlations such as the generative models or seed. 

\section{Domain generalization}
\label{sec:ood_generalization}
The new measurements from the lab inspire modifications in the candidate proposals or experimental protocol, which lead to (feedback) covariate or concept shifts. To account for the occurring shifts we turn to domain generalization.
DG has recently gained traction in the ML community as concerns about productionalizing ML models in unseen test environments have emerged \citep{rosenfeld2020risks}. 
The interest in achieving out-of-distribution (OOD) generalization has spawned a large body of work, which can be organized into the following families of approaches.

\textbf{DG by invariance.}
This paradigm is primarily motivated by learning ``causal/stable representations''. Invariant causal prediction (ICP) \citep{peters2016causal} frames prediction through causality, assuming data generation follows a structural equation model (SEM) where variables are related to their parents by structural equations. ICP assumes data partitioning into environments \( e \in E \) such that each environment corresponds to interventions on the SEM, but the mechanism generating the target variable via its direct parents remains unchanged \citep{pearl2009causality}. Thus, the causal mechanism of the target variable is fixed, while other generative distribution features can vary. This drives the goal of learning mechanisms that are stable across environments, hoping they generalize under unseen, valid interventions\footnote{Interventions are considered valid if they do not change the structural equation of \( Y \).}. The ultimate goal is to learn an ``optimal invariant predictor'' using only the invariant features of the SEM. We assume high-dimensional observations have lower-dimensional representations governed by a generative model. In the invariant learning paradigm, the task is commonly defined as learning invariant representations of the data, rather than invariant features in the observation space. This setup has inspired many algorithms, starting with IRM.\citep{arjovsky2019invariant} 
\begin{align*}
    \mathcal{R}_{\rm IRM} = \min_{\substack{\Phi: \mathcal{X}\rightarrow \mathcal{H}; \\ w:\mathcal{H} \rightarrow \mathcal{Y}}} \sum_{e \in E_{tr}} \mathcal{R}^e(w \cdot \Phi) ~s.t.~ w \in \underset{\bar{w}: H \rightarrow Y}{\arg\min} \ \mathcal{R}^e (\bar{w} \cdot \Phi) \ \forall e \in  E. 
\end{align*}
IRM assumes invariance of $\mathbb{E}[y \vert \Phi(x)]$---that is, invariance of the feature-conditioned label distribution. 
Follow-up studies make a stronger assumption on invariance based on higher-order conditional moments \citep{krueger2021out, xie2020risk}. 
Though this perspective has gained traction in the last few years, it is somewhat similar to methods from domain adaptation, such as DANN \citep{ganin2016domain} and CORAL \citep{sun2016deep}, which minimize domain shift by aligning the source and target feature distributions.
Another line of work considers learning shared mechanisms by imposing invariance of the gradients distributions across domains. In our setup the gradients per environment are $\mathbf{g}_{e} =  \mathbb{E}_{(X^e, Y^e)\sim \mathcal{D}_{e} }  \nabla_{\theta} \ell \left(\hat{f}_\theta(X^e), Y^e \right)$.
\citep{parascandolo2020learning} initiated such approaches, aiming to learn invariant explanations by replacing the arithmetic mean in gradient descent with a geometric one, hence promoting agreements of the gradients across domains. Other popular gradient based approaches include Fish \citep{shi2021gradient} which match the first moments of the gradient distributions, and Fishr \citep{rame2022fishr} which similarly to CORAL matches the variance in gradient space. 

\textbf{DG by ensembling.}
We consider two types of ensembling strategies that do not use domain labels.
First, \emph{output-space ensembles} combine multiple independently trained models for an input $\mathbf{x}$:
\begin{align*}
    \underset{k} {\rm arg max}~ \ {\rm Softmax}\left(\frac{1}{M}\sum_{m=1}^Mf(\mathbf{x}; \theta_m) \right)_k,
\end{align*}
where $M$ is the total number of models in the ensemble, $\theta_m$ are the parameters of the $m$-th model, and the sub-script $(\cdot)_k$ denotes the $k$-th element of the multiclass vector argument.
A standard ensembling approach, deep ensemble, combines models trained with different initializations and was shown to achieve strong robustness to OOD data \citep{lakshminarayanan2017simple}. These types of models will also be refereed to as 
 \emph{functional ensembles}.
Second, \emph{weight-space ensembles}.
Given $M$ individual member weights $\lbrace \theta_m\rbrace_{m=1}^M$ corresponding to individual models,  Weight averaging (WA), is defined as:
\begin{align*}
    f_{WA}=f(\cdot, \theta_{WA}), \text{where} \quad \theta_{WA} = \frac{1}{M}\sum_{m=1}^M \theta_{m}.
\end{align*}
A combination of different weight averaging and fine tuning resulted in different methods e.g. Stochastic Weight Average \citep{izmailov2018averaging}, Simple Moving Average \citep{arpit2022ensemble}, Diverse Weight Averaging a.k.a model soup \citep{wortsman2022model,rame2022diverse}.
These models usually leverage pretrained \emph{foundational} models \citep{bommasani2021opportunities}.

\textbf{Hypothesis: invariant feature representations of antibodies.}
Our lab-in-the-loop (\autoref{sec:lll}) offers a unique testbed for DG algorithms. In particular, we attempt to answer the question:

\emph{
Can DG algorithms help in developing robust predictors for antibody design? Do learnt invariant representations align with the physics-based features causing binding properties?
}

We propose to consider the design rounds $r \in \lbrace 0, \dotsc, 4 \rbrace$ as environments $e$, since rounds do correspond to valid interventions --- our design cycles should not impact the true causal mechanism governing binding affinity. 
There are two types of features that a binding classifier can learn:
\begin{tightemize}
    \item \emph{Invariant (causal) features}: various physico-chemical and geometric properties at the interface of antibody-antigen binding (\autoref{fig:context}) and
    \item \emph{Spurious correlations}: Other round-specific features that are byproducts of different folding algorithms, generative models, measurement assay types, antigen targets, etc. 
\end{tightemize}
We expect DG algorithms to be able to distinguish between the two, and only make use of the features invariant across rounds in their predictions. 
\section{Related work}
\label{sec:related_work}
Existing OOD generalization benchmarks are predominantly image-based \citep[e.g.,][]{gulrajani2020search,koh21a}, and it remains unclear if their conclusions apply to real-world applications. Our benchmark addresses this gap, focusing on drug discovery.
Previous benchmarks in drug discovery have been limited to small molecules \citep{Zhang2023, Tossou2023}. In contrast, we propose a benchmark for therapeutic antibodies, considering distribution shifts typical in active ML-guided design. Antibodies, unlike small molecules, are large, complex structures made of thousands of atoms, offering high target selectivity and reduced toxicity \citep{shepard2017developments}. Their three-dimensional folded structure is crucial for predicting functional properties, making structure-aware representations essential.

\citep{minot2024meta} and \citep{hummer2023investigating} are the only other public datasets focusing on antibodies. The former addresses noise in high-throughput experimental measurements of affinity, which impacts ML predictors, while the latter highlights structure-based prediction challenges under train-test discrepancy, without proposing solutions beyond data collection. Our work complements these by focusing on various sources of distribution shifts accompanying active design.
Proteins pose unique modeling challenges, and, to our knowledge, this is the first large-scale OOD benchmark on large molecules that considers sequence, structure, and language-based foundational model representations for property prediction.
\section{Antibody DomainBed}
\label{sec:domain_bed}

\subsection{Therapeutic protein dataset.}
The primary goal of this benchmark is to emulate a real-world active drug design setup. To create a realistic, publicly accessible dataset, we propose the following steps: (1) Collect open data on antibody-antigen complex structures; 
(2) Train generative models to sample antibodies with varying statistics (e.g., edit distances from training data, targets of interest, different seed complexes); (3) Compute binding proxies using physics-based models for all designs from (2); and (4) Split the labeled dataset into meaningful environments for application of DG algorithms. In \autoref{sec:app_ab_valid}, we show that our antibodies span reasonable ranges in various metrics including biophysical properties (e.g., hydrophobicity, aromaticity) and distributional scores evaluating proximity to a database of natural antibodies (e.g., naturalness \citep{olsen2022observed}).

\textbf{Step 1: Data curation.} We select antibodies associated with the antigens HIV1\footnote{Most common type of HIV that can lead to AIDS. HIV attacks the body's immune system by destroying CD4 cells, which help your body fight infections.}, SARS‑CoV‑2,\footnote{Severe acute respiratory syndrome coronavirus 2, is a strain of coronavirus that causes COVID-19.} and HER2 \footnote{Human epidermal growth factor receptor 2 is a gene that makes a protein found on the surface of all breast cells. Breast cancer cells with higher levels of HER2 signal breast cancer may grow quickly and possibly recur.}. For Env 0-3, we post-process paired antibody-antigen structures in the latest version (released in 2022) of the \emph{Structural Antibody Database}, SAbDab \cite{dunbar2014sabdab, raybould2020thera, schneider2022sabdab}. Antibody sequences corresponding to 179 SAbDab structures for the three antigens serve as the starting seeds for our sequence-based generative models (see Step 2), and we modify the structures based on mutations made by the model to compute the labels (see Step 3). For Env 4, we take a recent derivative \emph{Graphinity}, \cite{Hummer2023} which extends SAbDab to a synthetic dataset of a much larger scale ($\sim$1M) by introducing systematic point mutations in the CDR3 loops of the antibodies in the SAbDab complexes. 

\textbf{Step 2: Sampling antibody candidates.} To emulate the active drug discovery pipeline, we need a suite of generative models for sampling new candidate designs for therapeutic antibodies. We run the \emph{Walk Jump Sampler} \citep[WJS;][]{frey2023protein}, a method building on the neural empirical Bayes framework \citep{saremi2019neural}.
WJS separately trains score- and energy-based models to learn noisy data distributions and sample discrete data. The energy-based model is trained on noisy samples, which means that by training with different noise levels $\sigma$, we obtain different generative models. Higher $\sigma$ corresponds to greater diversity in the samples and higher distances from the starting seed. We used four values for the noise parameter, namely $\sigma \in \{0.5, 1.0, 1.5, 2.0\}$. 

\textbf{Step 3: Labeling candidates.}
As lab assays to experimentally measure binding affinity are prohibitively expensive, we use computational frameworks which, by modeling changes in binding free energies upon mutation (interface $\Delta\Delta G = \Delta G_\textrm{wild type} - \Delta G_{\rm mutant}$), enable large-scale prediction and perturbation of protein–protein interactions \cite{barlow2018flex}. We use the \verb|pyrosetta| (pyR)  \cite{chaudhury2010pyrosetta} implementation of the \verb|InterfaceAnalyzerMover|, namely the scoring function \verb |ref2015|, to compute the difference in energies before and after mutations in a given antibody-antigen complex. After removing highly uncertain labels between -0.1 and 0.1 kcal/mol \citep{sirin2016ab,Hummer2023}, we attach binary labels to each candidate of the generative models: label 1 if $\Delta\Delta G < -0.1$ (stabilizing) and label 0 if $\Delta\Delta G > 0.1$ (destabilizing). While computed $\Delta\Delta G$ represent weak proxies of binding free energy, they have been shown to be predictive of experimental binding \citep{mahajan2022hallucinating,Hummer2023, faure2022mapping}. See \ref{app:ddg} for details.

\textbf{Step 4: Splitting into environments.}
To emulate the active drug design setup where distribution shifts may appear in each design cycle (e.g., due to new generative models, antigen targets, or experimental assays), we split the overall dataset into five environments. In \autoref{tab:ds} we summarize our environment definitions as well as key summary statistics of each environment. For Env 0-3, the dataset split mimics sub-population shifts due to the generative model, which produces antibody sequences with different edit distances from the seed sequences. The WJS model with $\sigma{\rm =}0.5$ ($\sigma{\rm =}2.0$) produces antibody designs close to (far from) the seed. Env 4 has been partially included in the experiments because it introduces severe distribution shift in the form of concept drift and label shift, as it represents a completely new target and a different labeling mechanism than the rest. We report preliminary results in this extreme setup in \autoref{sec:app_her2}. 
\begin{table}
\centering
\caption{Dataset overview.}\label{tab:ds}
\resizebox{\textwidth}{!}{
\begin{tabular}[width=0.9\textwidth]{@{}lllll@{}}
\toprule
\textbf{Environment}      & \textbf{Antigens (number of samples)}                                           & \textbf{Generative model}   & \textbf{$\Delta \Delta G$ annotation} & \textbf{Type of distribution shift emulated}              \\ \midrule
\textbf{Env 0} & \begin{tabular}[c]{@{}l@{}}HIV1 (186)\\ SARS‑CoV‑2 (1117)\end{tabular} & WJS $\sigma \in [0.5]$ & pyRosetta &  covariate shift 
\\
\midrule
\textbf{Env 1} & \begin{tabular}[c]{@{}l@{}}HIV1 (780)\\ SARS‑CoV‑2 (3096)\end{tabular} & WJS $\sigma \in [1.0, 1.5]$ & pyRosetta & covariate shift due 
\\
\midrule
\textbf{Env 2} & \begin{tabular}[c]{@{}l@{}}HIV1 (275)\\ SARS‑CoV‑2 (1469)\end{tabular} & WJS $\sigma \in [2.0]$ & pyRosetta  & covariate shift 
\\
\midrule
\textbf{Env 3} & \begin{tabular}[c]{@{}l@{}}HIV1 (552)\\ SARS‑CoV‑2 (3141)\end{tabular} & WJS $\sigma \in [0.5, 1.0, 1.5, 2.0]$&  pyRosetta & covariate shift due 
\\
\midrule
\textbf{Env 4$^*$} & HER2 (2434)&  point mutations in CDR3 & FoldX & zero-shot (new target), concept drift \\ 
\bottomrule
\end{tabular}
}
\end{table}
\vspace{-1em}
\begin{figure}%
    \centering
    \subfigure{\includegraphics[width=0.45\textwidth]{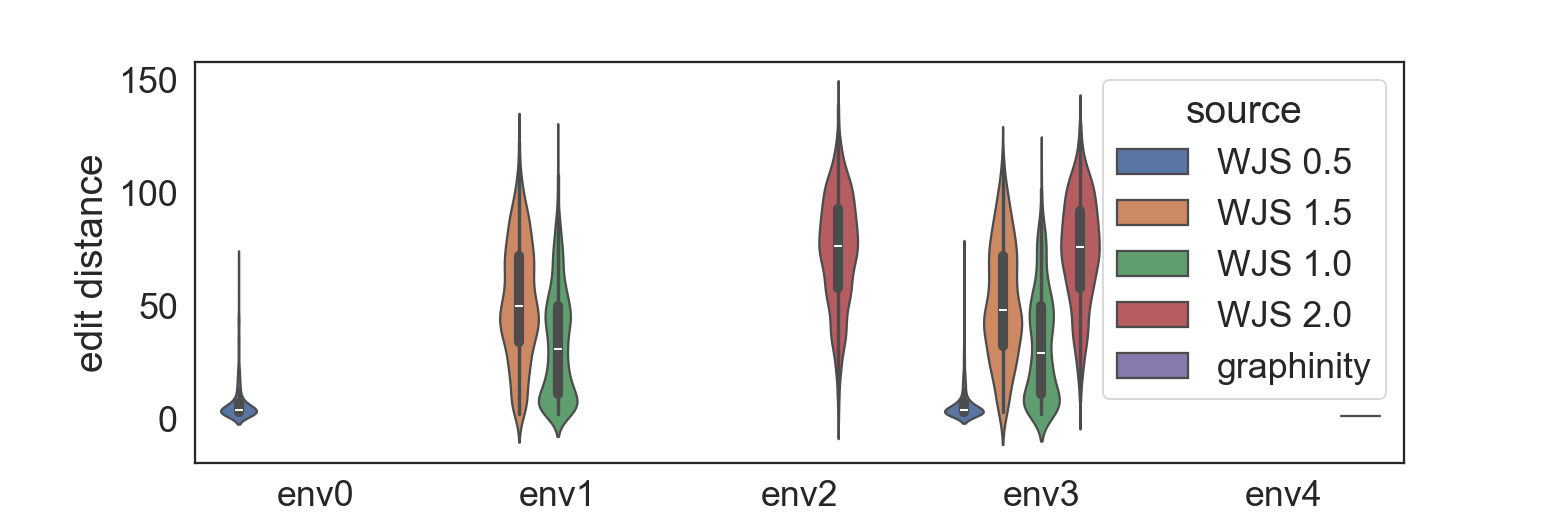}}%
    \subfigure{\includegraphics[width=0.55\textwidth]{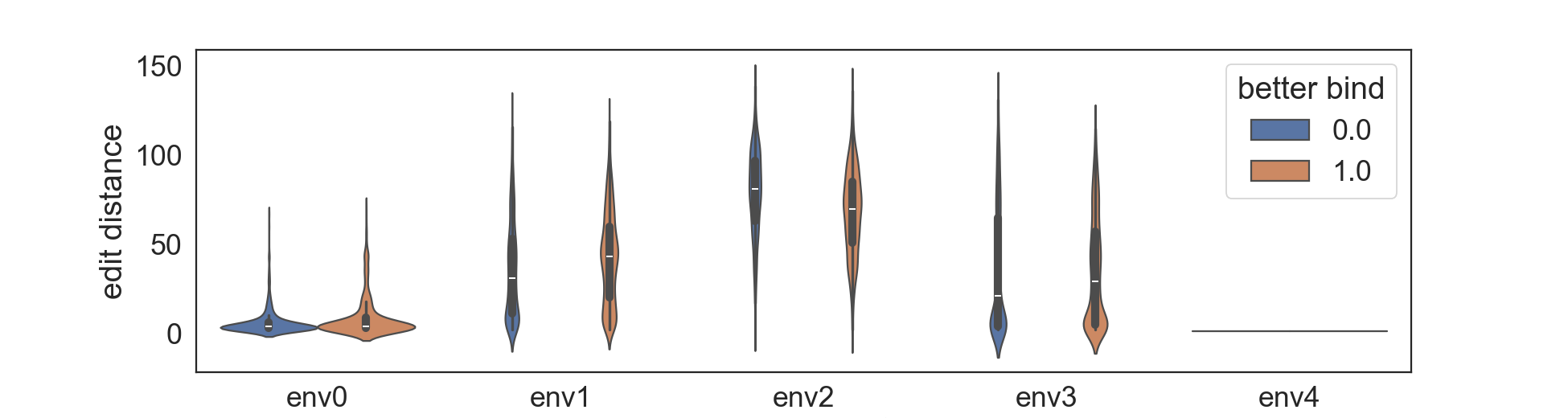} }%
    \caption{Antibody Domainbed environments. Left: edit distance (sequence similarity) between designs and seeds. Right: binding properties per generative model.}%
    \label{fig:example}%
\end{figure}
\vspace{-0.5em}

\subsection{Protein sequence benchmark}

DG algorithms vary in their assumptions of invariance and methods of estimating it from data. DomainBed \citep{gulrajani2020search} is a benchmark suite that implements DG algorithms developed in the past four years and provides a benchmark environment that compares them on multiple natural image datasets. 

To adapt DomainBed to our antibody design context, we modify its featurizer to accept biological sequences as input. We do so by (i) implementing a preprocessing module to align the antibody sequences \cite[using the AHo numbering scheme suitable for antibody variable domains;][]{honegger2001yet} and (ii) replacing the default ResNet \citep{he2016deep} with one of the following more suitable architectures:  

\textbf{SeqCNN:} Our SeqCNN model consists of an embedding layer with output dimension 64, and two consecutive convolutional layers  with kernel sizes 10 and 5 respectively, stride of 2 followed by ReLU nonlinearities. The CNN output is pooled with a mixing layer of size 256. This identical architecture is applied for both the antibody and the antigen sequence before concatenating them and passing them to the classification head.  

\textbf{Finetuned ESM2:} We finetune the 8M-parameter ESM2 \cite{lin2023evolutionary}, a protein language model pretrained on experimental and high-quality predicted structures of general proteins. 
We used a single ESM2 model for the two antibody chains as well as the antigen.

\textbf{GearNet:} {We finetune GearNet-Edge MVC \citep{zhang2022protein}, a general-purpose structure-based protein encoder pretrained with a contrastive learning objective on AlphaFold2 \citep{jumper2021highly} predictions. Notably, the pretraining dataset only has monomers and does not include a significant antibody fraction. As a result, the Antibody DomainBed environments are OOD for GearNet-ESM. We make no adjustments, using the default structure graph construction and featurization.}

Additionally, we extend DomainBed to include the moving average ensembling as in \citep{arpit2022ensemble} as well as {functional ensembling and simple weight averaging \citep{arpit2022ensemble} for all DG baselines in the repository. We denote the averaged solutions with \emph{-ENS} suffix.}

\begin{figure}
    \centering
     \includegraphics[trim={8cm 3.5cm 3.5cm 4.2cm}, clip, width=0.9\textwidth]{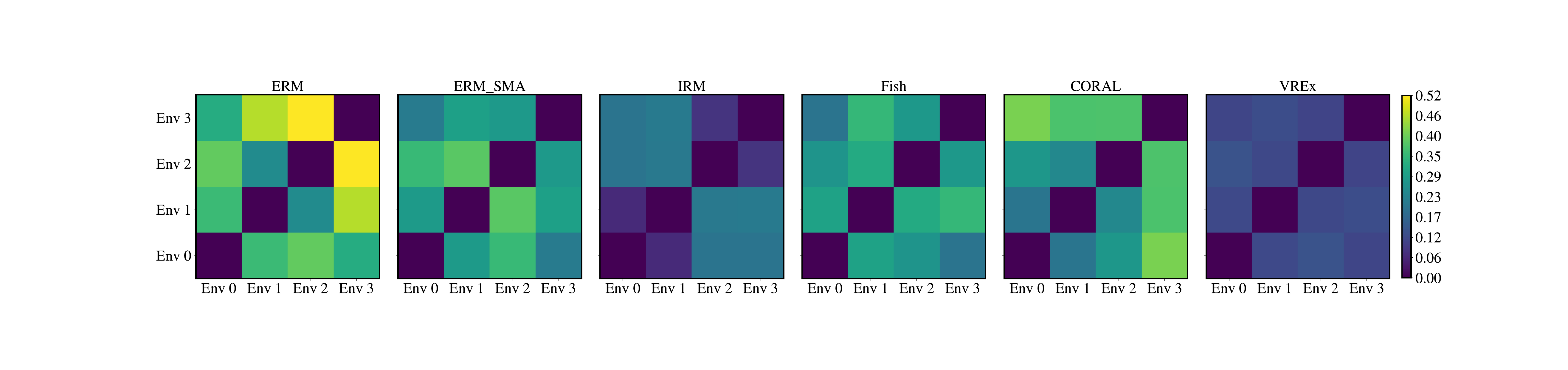}
\caption{MMD in the learned features of ESM between every pair of environments. DG algorithms result in features that are significantly more uniform across environments.\label{fig:mmd}}
\vspace{-0.5cm}
 \end{figure}

\textbf{Code and data sharing.}
We open-source our work to enable further research on similar biological data scenarios. This includes the ``Antibody DomainBed'' codebase, aligned with the DomainBed suite, available \href{https://github.com/prescient-design/antibody-domainbed}{here \faGithub}, and a public benchmark dataset of paired antibody-antigen sequences and structures, available \href{https://www.dropbox.com/scl/fo/e670i9adp29yv2knfu6wd/h?rlkey=uax6phjjfumkk8xoxrbwcit1h&dl=0}{here}.

\section{Benchmarking Results and Analysis}
\label{sec:results}
We tested vanilla ERM against three classes of DG methods.
These are methods that leverage the domain information to enforce invariance on representations \citep[CORAL;][]{sun2016deep}, on the predictor \citep[IRM;][]{arjovsky2019invariant}, or on gradients \citep[Fish;][]{shi2021gradient}.
We additionally test ensembling techniques, namely deep ensembles, \citep[simple moving average or SMA;][]{arpit2022ensemble}, that do not exploit domain information but combine different models to reduce the effect of covariate shift \citep{rame2022diverse}. 
The last class of models has achieved state-of-the-art results on image datasets \citep{arpit2022ensemble,rame2022diverse,cha2021swad}.
We perform early stopping for the members of the deep ensembles in the training-domain validation setting as in \cite{arpit2022ensemble}.
The best model is ensembled based on validation performance over each trial.
From the available DG algorithms in DomainBed, this corresponds to 6 algorithm baselines with 5 hyperparameter configurations (with varying batch size, weight decay, and learning rate) for ESM-based architectures and 20 for SeqCNN-based architectures. As we perform 3 seed repetitions for each configuration, we have a total of 1,890 experiments. We report the results from \emph{model selection method: training domain validation set}, which is a leave-one-environment-out model selection strategy. We (1) split the data into train and test environments, (2) pool the validation sets of each training domain to create an overall validation set, and (3) choose the model maximizing accuracy (minimizing the negative log-likelihood) on the pooled validation set.



\textbf{Metrics and evaluation.}
As in \cite{gulrajani2020search} we evaluate the classification accuracy in two types of validation: training-domain and test-domain.
We also compute saliency maps from a representative subset of domain-invariant and ensembling models. In \autoref{sec:app_her2}, we repeat the analysis for a second version (a new split) of the dataset containing an environment with antibody designs of a new antigen.

\textbf{Results}
\looseness-1 Based on the results in \autoref{tab:train_res} and \autoref{tab:test_res} (in \autoref{sec:app_results})  we observe that, 
among the different DG paradigms, ensembling-based models work the best and functional ensembles improve all base algorithms. Additionally, as the model size increases (from SeqCNN to ESM2), the improvement is more pronounced. 
When the validation set is not aligned with the test set, there is no significant difference between models in terms of performance.
The only class of models robust to such a setup is, again, functional ensembles; they seem to work well even if the validation set is not aligned to the test set.

SMA is the only DG model that has advantage over ERM across all domains. From the maximum mean discrepancy (MMD) plots in \autoref{fig:mmd}, we do notice that the invariance based models indeed learn more uniform representations across domains. It seems, however, that invariant representations do not necessary translate to better predictive performance. 


\begin{figure}[th]
    \centering
    \begin{subfigure}{\includegraphics[trim={0cm 0.5cm 0cm 2.25cm}, clip, width=0.32\textwidth]{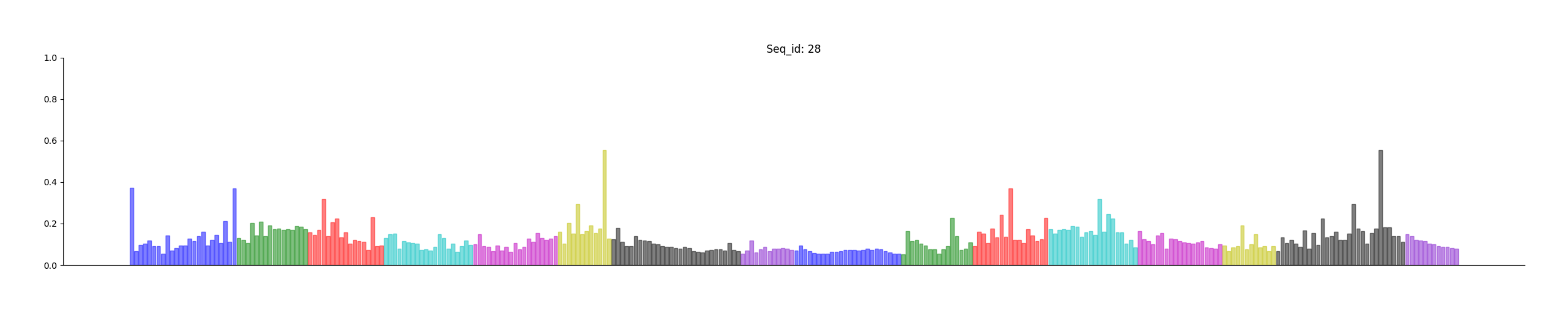}}
     \end{subfigure}
     \begin{subfigure}{\includegraphics[trim={0cm 0.5cm 0cm 2.25cm}, clip, width=0.32\textwidth]{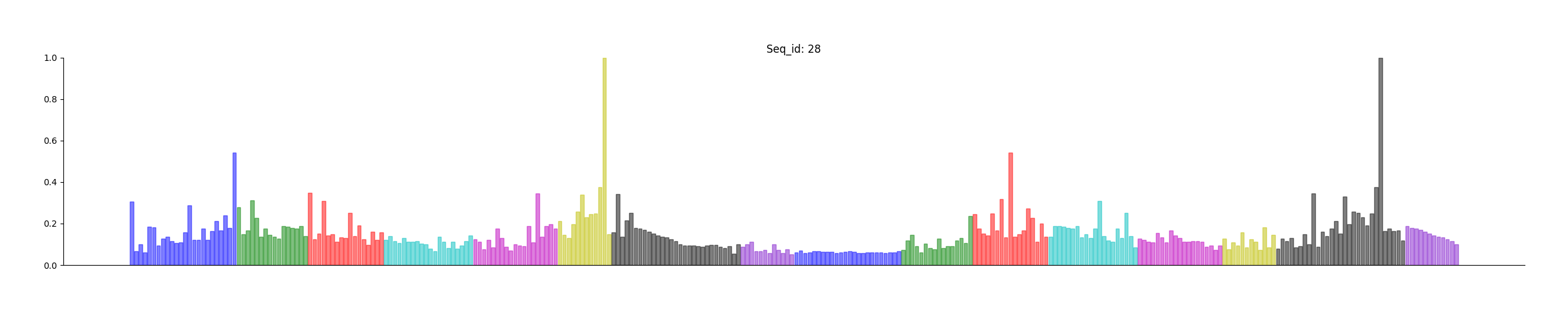}}
     \end{subfigure}
     \begin{subfigure}{\includegraphics[trim={0cm 0.5cm 0cm 2.25cm}, clip, width=0.32\textwidth]{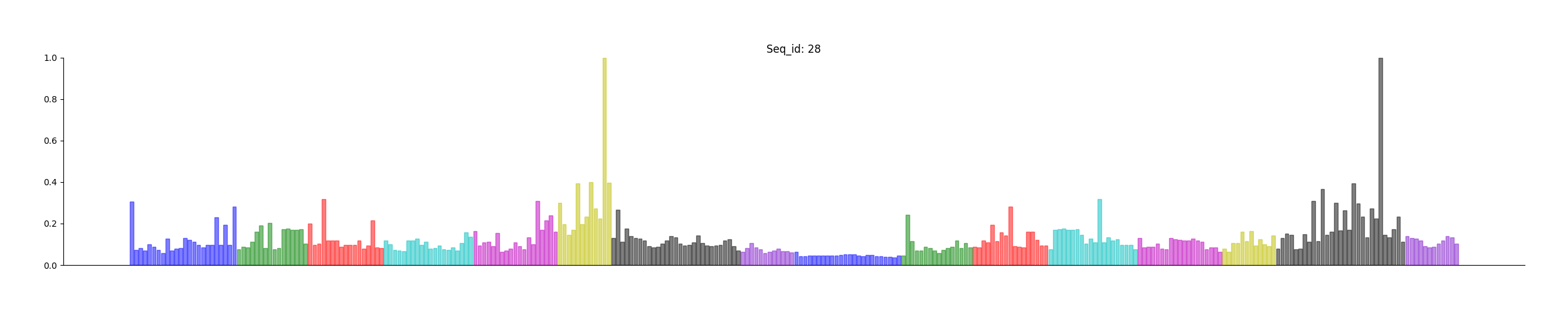}}
     \end{subfigure}
\caption{Saliency visualizations for ERM, SMA, and IRM on the antibody residue positions for heavy and light chains.
Colors of bars represent functional segments. Relative to ERM-SMA and IRM, ERM displays muted behavior in the regions known to interact with the antigen paratope. \label{fig:saliency}}
 \end{figure}

\begin{wrapfigure}{r}{0.45\textwidth}
    \small
    \begin{center}
\includegraphics[width=0.4\textwidth]{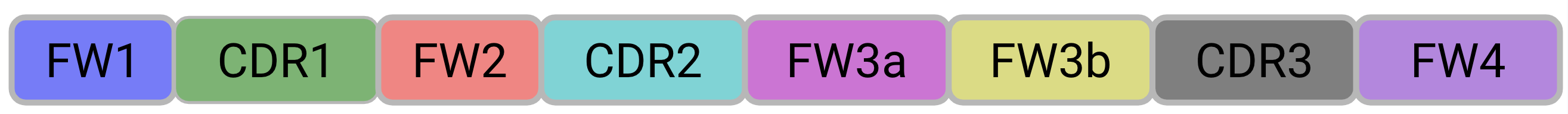} 
\end{center}
\caption{Functional segments of the variable (Fv) region of an antibody.} \label{fig:legend}
\vspace{-0.2cm}
\end{wrapfigure} 
 
\autoref{fig:saliency} shows the gradient sensitivity \citep{baehrens2010explain,simonyan2013deep} along the antibody residue positions. Colors represent functional segments, labeled in \autoref{fig:legend}. Details on the structure and function of the different segments in an antibody are included in \autoref{sec:ab_str}. Spikes can be interpreted as positions that strongly affected the prediction. Whereas ERM saliency is muted in the right edges of framework 3b (yellow) in the heavy chain and in CDR 3 (gray) in the light chain, ERM-SMA and IRM are activated in these regions. Heavy framework 3b, adjacent to heavy CDR 3, is heavily implicated in antigen binding \citep{morea1998conformations}. Light CDR 3 displays high diversity, though to a lesser extent than the heavy counterpart. It is known to assume one of the few canonical structures that determines antigen recognition \citep{teplyakov2014canonical}, which may explain the models' sensitivity to this segment.

 Our main takeaways for robust prediction in therapeutic protein design are as follows:
\begin{tightemize}
        \item Leveraging foundational models pretrained on large protein databases and ensembling multiple models help performance across all DG models.
    \item Keeping a consistent validation set is important. This may be achieved, for instance, by fixing the set of seeds across design rounds to control for covariate shift.
    \item Structure-based models outperform the ESM-based model and SeqCNN in $\Delta \Delta G$ prediction, though at the price of more compute. Their DG variants require careful parameter tuning.
\end{tightemize}

\begin{table}[ht]
\begin{center}
 \caption{Model selection: training-domain validation set. Best in bold. Second best underlined. \label{tab:train_res}} 
 \adjustbox{width=0.7\textwidth}{%
 \small
        \begin{tabular}{lccccc}
        \toprule
        \textbf{Algorithm}   & \textbf{Env 0}     & \textbf{Env 1}     & \textbf{Env 2}     & \textbf{Env 3}     & \textbf{Avg}         \\
        \midrule
         Random                   & 50.5 $\pm$ 0.0       & 52.4 $\pm$ 5.2       & 38.8 $\pm$ 0.0       & 49.1 $\pm$ 1.1       & 47.7                 \\
         \midrule
        \multicolumn{6}{c}{ESM: foundational sequence-based architecture} \\
        \midrule
        ERM                  & 58.7 $\pm$ 0.7      & 66.8 $\pm$ 0.4      & 69.0 $\pm$ 0.5       & 65.9 $\pm$ 0.2       & 65.1                 \\
        ERM-ENS                  & 62.9       & 69.7      & 71.6       & 67.9       &   68.0              \\
        SMA              & 59.6 $\pm$ 1.1       & 66.7 $\pm$ 0.0       & 70.1 $\pm$ 0.1       & 66.2 $\pm$ 0.3       & 65.7                 \\
        SMA-ENS                   & 61.5      & 68.1      &  73.4      &   67.7     &  67.7               \\
        IRM                  & 60.0 $\pm$ 0.9       & 64.4 $\pm$ 0.2       & 69.6 $\pm$ 0.9       & 63.5 $\pm$ 0.6       & 64.4                 \\
        IRM-ENS                  &  62.7     & 69.1       & 71.3       & 67.0       & 67.5                \\
        CORAL                & 60.0 $\pm$ 0.4       & 66.9 $\pm$ 0.2       & 69.6 $\pm$ 0.5       & 65.7 $\pm$ 0.4       & 65.5                 \\
        CORAL-ENS                  & 62.6       & 69.4       & 71.7       & 68.1       & 68.0                 \\
        VREx                 & 58.7 $\pm$ 0.6       & 66.1 $\pm$ 0.9       & 69.0 $\pm$ 0.6       & 65.7 $\pm$ 0.1       & 64.9                    \\
        VREx-ENS             & 61.1       & 67.2       &  71.9      & 67.9       & 67.0                \\
        Fish                 & 59.3 $\pm$ 1.2       & 66.2 $\pm$ 0.7       & 69.5 $\pm$ 0.1       & 66.4 $\pm$ 0.4       &  65.3                    \\
        \midrule
 \multicolumn{6}{c}{SeqCNN: sequence-based convolutional architecture} \\
 \midrule
 ERM                  & \underline{63.2 $\pm$ 1.0 }     & 66.2 $\pm$ 0.9       & 66.2 $\pm$ 0.8       & 64.9 $\pm$ 0.1       & 65.1                 \\
ERM-ENS   &  62.3      &          66.1          &    69.7             &  65.9 & 66.0\\
SMA              & 61.8 $\pm$ 0.9       & 66.5 $\pm$ 0.3       & 66.1 $\pm$ 0.2       & 64.9 $\pm$ 0.3       & 64.9                 \\             
SMA ENS             & 58.6    &   66.9     & 68.8      & 66.1    & 65.0                 \\
IRM                  & 60.0 $\pm$ 0.9       & 64.4 $\pm$ 0.2       & 69.6 $\pm$ 0.9       & 63.5 $\pm$ 0.6       & 64.4                 \\
IRM-ENS                  & 62.4      &  66.5      & 73.2     & 65.1      &   66.8               \\
CORAL                & 62.0 $\pm$ 1.5       & 66.2 $\pm$ 0.3       & 66.1 $\pm$ 0.8       & 64.0 $\pm$ 0.5       & 64.6                 \\
CORAL-ENS                & 60.9        & 67.1       & 68.9       & 65.2       & 65.5                 \\
VREx                 & 60.1 $\pm$ 1.6       & 65.7 $\pm$ 1.0       & 66.3 $\pm$ 0.6       & 64.9 $\pm$ 0.4       & 64.2                 \\
VREx-ENS                            & 61.5      & 66.9       & 68.2      & 66.1     & 65.7                 \\
Fish                 & 62.0 $\pm$ 0.9       & 66.9 $\pm$ 0.8       & 68.6 $\pm$ 0.6       & 65.8 $\pm$ 0.3       & 65.8   \\
\midrule
 \multicolumn{6}{c}{GearNet: foundational structure-based graph architecture} \\
\midrule
ERM &                 60.4 $\pm$ 0.6       & \underline{79.2 $\pm$ 0.3}       & 85.3 $\pm$ 0.5       & \underline{75.1 $\pm$ 1.1}       & 75.0    \\
ERM-ENS               & \textbf{64.9}      & \textbf{82.0}       & \textbf{88.0}       &  \textbf{78.4}      & \textbf{78.3}                 \\
VREx                 & 58.3 $\pm$ 0.4       & 73.1 $\pm$ 0.5       & 82.1 $\pm$ 1.3       & 71.1 $\pm$ 1.3       & 71.2                 \\
VREx-ENS                 & 59.5       & 77.6 & \underline{86.4}      & 73.3       & 74.2                     \\
\bottomrule
\end{tabular}}
\end{center}
\vspace{-0.2cm}
\end{table}

\begin{wraptable}{r}{0.45\textwidth}
    \small
    \begin{center}
    \begin{tabular}[width=0.45\textwidth]{@{}lcc@{}}
    \toprule
    & \textbf{Train acc} & \textbf{Test acc} \\ \midrule
    \textbf{Random} & 49.0              & 50.0              \\
    \textbf{Metadata ERM}     & 60.0       & 59.0     \\
    \textbf{SeqCNN ERM}        & 64.0               & 65.1        \\ \bottomrule
    \end{tabular}
\end{center}
\caption{An ERM model trained only on metadata performs similarly to an ERM model trained on sequences (SeqCNN).} \label{fig:metadata}
\vspace{-0.2cm}
\end{wraptable}
\textbf{Metadata experiment.}
We investigate the possibility that the model uses spurious features present in various metadata that can be easily inferred from the sequences. For instance, the identity of the target antigen can be mapped directly from the input antigen sequence. Similarly, generative models leave identifiable signatures on their designs, so the type of generative model can be inferred from sequence information. Latching onto such metadata, the model can make shortcut predictions based mainly on the combination of the source algorithm and the target antigen (e.g., predict positive if the antibody sequence originated from a generative model that output a high proportion of positives in previous rounds). To investigate this hypothesis, we train a tabular model using ERM with only the metadata: name of the generative model, name of the target, environment, and edit distance from seed. We compare its performance to SeqCNN, which takes the antigen and antibody sequences as inputs on an IID split. 

Table~\ref{fig:metadata} shows that simple tabular features are enough increase the test accuracy of the metadata model by 9\% relative to a random classifier. Moreover, the metadata model is worse by only 5\% compared to the model trained on the full sequence (SeqCNN). For ERM, there is no mechanism to deter the model from soaking in such spurious correlations.

\textbf{Limitations.} Benchmarking in Antibody DomainBed consumes significant resources, as it involves repeating the runs multiple times for every combination of environment, DG algorithm, hyperparameter configuration, and seed. While the structure-based predictors perform the best across ERM and DG settings, they impose additional complexity and take even longer to run (up to a day per run).

\section{Conclusion}
We publicly release the codebase of our Antibody DomainBed pipeline as well as the associated dataset of paired antibody-antigen sequences and structures. Representing the first large-molecule benchmark for OOD generalization, our experiments suggest that DG methods have the capacity to significantly aid protein property prediction in the presence of complex distribution shifts. Antibody DomainBed enables the exploration of key theoretical questions that, when addressed, would maximize the impact of DG methods on biological problems. One question is: how can we define environments that would lead to optimal performance at test time? Related to this is the question of how to choose the different configurations governing each environment in a manner that would maximize learning for each DG algorithm. Finally, the ultimate OOD quest in the context of antibody design would be to produce accurate predictions for a completely new antigen. This will require the DG models to pick up on truly causal, or mechanistic, features governing the binding interaction. By open sourcing our code as well as the data and generators, we hope to motivate other ML researchers to aim at addressing impactful real-world applications close to the production setting. 



\newpage
\bibliography{neurips_data_2024}

\begin{thebibliography}{10}
\expandafter\ifx\csname url\endcsname\relax
  \def\url#1{\texttt{#1}}\fi
\expandafter\ifx\csname urlprefix\endcsname\relax\def\urlprefix{URL }\fi
\expandafter\ifx\csname href\endcsname\relax
  \def\href#1#2{#2} \def\path#1{#1}\fi

\bibitem{torralba2011unbiased}
A.~Torralba, A.~A. Efros, Unbiased look at dataset bias, CVPR 2011 (2011).

\bibitem{zech2018variable}
J.~R. Zech, M.~A. Badgeley, M.~Liu, A.~B. Costa, J.~J. Titano, E.~K. Oermann, Variable generalization performance of a deep learning model to detect pneumonia in chest radiographs: a cross-sectional study, PLoS medicine 15 (2018).

\bibitem{beery2019efficient}
S.~Beery, D.~Morris, S.~Yang, Efficient pipeline for camera trap image review, Data Mining and AI for Conservation Workshop at KDD19 (2019).

\bibitem{koh2021wilds}
P.~W. Koh, S.~Sagawa, H.~Marklund, S.~M. Xie, M.~Zhang, A.~Balsubramani, W.~Hu, M.~Yasunaga, R.~L. Phillips, I.~Gao, et~al., Wilds: A benchmark of in-the-wild distribution shifts, ICML (2021).

\bibitem{neuhaus2022spurious}
Y.~Neuhaus, M.~Augustin, V.~Boreiko, M.~Hein, Spurious features everywhere--large-scale detection of harmful spurious features in imagenet, ICCV (2023).

\bibitem{blanchard2011generalizing}
G.~Blanchard, G.~Lee, C.~Scott, Generalizing from several related classification tasks to a new unlabeled sample, NeurIPS 24 (2011).

\bibitem{muandet2013domain}
K.~Muandet, D.~Balduzzi, B.~Sch{\"o}lkopf, Domain generalization via invariant feature representation, ICML (2013).

\bibitem{arjovsky2019invariant}
M.~Arjovsky, L.~Bottou, I.~Gulrajani, D.~Lopez-Paz, Invariant risk minimization, arXiv preprint arXiv:1907.02893 (2019).

\bibitem{ahuja2021invariance}
K.~Ahuja, E.~Caballero, D.~Zhang, J.-C. Gagnon-Audet, Y.~Bengio, I.~Mitliagkas, I.~Rish, Invariance principle meets information bottleneck for out-of-distribution generalization, NeurIPS (2021).

\bibitem{rame2022fishr}
A.~Rame, C.~Dancette, M.~Cord, Fishr: Invariant gradient variances for out-of-distribution generalization, ICML (2022).

\bibitem{gulrajani2020search}
I.~Gulrajani, D.~Lopez-Paz, In search of lost domain generalization, ICLR (2021).

\bibitem{lynch2023spawrious}
A.~Lynch, G.~J. Dovonon, J.~Kaddour, R.~Silva, Spawrious: A benchmark for fine control of spurious correlation biases, CVPR (2023).

\bibitem{singh2018monoclonal}
S.~Singh, N.~K. Tank, P.~Dwiwedi, J.~Charan, R.~Kaur, P.~Sidhu, V.~K. Chugh, Monoclonal antibodies: a review, Current clinical pharmacology 13 (2018).

\bibitem{rolfo2015onartuzumab}
C.~Rolfo, N.~Van Der~Steen, P.~Pauwels, F.~Cappuzzo, Onartuzumab in lung cancer: the fall of icarus? (2015).

\bibitem{gligorijevic2021function}
V.~Gligorijevi{\'c}, D.~Berenberg, S.~Ra, A.~Watkins, S.~Kelow, K.~Cho, R.~Bonneau, Function-guided protein design by deep manifold sampling, Machine Learning for Structural Biology Workshop, NeurIPS (2021).

\bibitem{berenberg2022multi}
D.~Berenberg, J.~H. Lee, S.~Kelow, J.~W. Park, A.~Watkins, V.~Gligorijevi{\'c}, R.~Bonneau, S.~Ra, K.~Cho, Multi-segment preserving sampling for deep manifold sampler, Machine Learning for Drug Discovery Workshop, ICLR (2022).

\bibitem{tagasovska2022pareto}
N.~Tagasovska, N.~C. Frey, A.~Loukas, I.~H{\"o}tzel, J.~Lafrance-Vanasse, R.~L. Kelly, Y.~Wu, A.~Rajpal, R.~Bonneau, K.~Cho, et~al., A pareto-optimal compositional energy-based model for sampling and optimization of protein sequences, NeurIPS AI4Science Workshop (2022).

\bibitem{frey2023learning}
N.~C. Frey, D.~Berenberg, J.~Kleinhenz, I.~Hotzel, J.~Lafrance-Vanasse, R.~L. Kelly, Y.~Wu, A.~Rajpal, S.~Ra, R.~Bonneau, et~al., Learning protein family manifolds with smoothed energy-based models, ICLR 2023 Workshop on Physics for Machine Learning (2023).

\bibitem{gruver2023protein}
N.~Gruver, S.~Stanton, N.~C. Frey, T.~G. Rudner, I.~Hotzel, J.~Lafrance-Vanasse, A.~Rajpal, K.~Cho, A.~G. Wilson, Protein design with guided discrete diffusion, NeurIPS (2023).

\bibitem{frey2023protein}
N.~C. Frey, D.~Berenberg, K.~Zadorozhny, J.~Kleinhenz, J.~Lafrance-Vanasse, I.~Hotzel, Y.~Wu, S.~Ra, R.~Bonneau, K.~Cho, et~al., Protein discovery with discrete walk-jump sampling, ICLR (2024).

\bibitem{park2022propertydag}
J.~W. Park, S.~Stanton, S.~Saremi, A.~Watkins, H.~Dwyer, V.~Gligorijevic, R.~Bonneau, S.~Ra, K.~Cho, Propertydag: Multi-objective bayesian optimization of partially ordered, mixed-variable properties for biological sequence design, AI4Science Workshop at NeurIPS (2022).

\bibitem{vapnik1992principles}
V.~Vapnik, Principles of risk minimization for learning theory (1992).

\bibitem{rosenfeld2020risks}
E.~Rosenfeld, P.~Ravikumar, A.~Risteski, The risks of invariant risk minimization, ICLR (2021).

\bibitem{peters2016causal}
J.~Peters, P.~B{\"u}hlmann, N.~Meinshausen, Causal inference by using invariant prediction: identification and confidence intervals, Journal of the Royal Statistical Society. Series B (Statistical Methodology) (2016).

\bibitem{pearl2009causality}
J.~Pearl, Causality, Cambridge university press, 2009.

\bibitem{krueger2021out}
D.~Krueger, E.~Caballero, J.-H. Jacobsen, A.~Zhang, J.~Binas, D.~Zhang, R.~Le~Priol, A.~Courville, Out-of-distribution generalization via risk extrapolation (rex), ICML (2021).

\bibitem{xie2020risk}
C.~Xie, F.~Chen, Y.~Liu, Z.~Li, Risk variance penalization: From distributional robustness to causality, arXiv preprint arXiv:2006.07544 (2020).

\bibitem{ganin2016domain}
Y.~Ganin, E.~Ustinova, H.~Ajakan, P.~Germain, H.~Larochelle, F.~Laviolette, M.~Marchand, V.~Lempitsky, Domain-adversarial training of neural networks, JLMR 17 (2016).

\bibitem{sun2016deep}
B.~Sun, K.~Saenko, Deep coral: Correlation alignment for deep domain adaptation, ECCV (2016).

\bibitem{parascandolo2020learning}
G.~Parascandolo, A.~Neitz, A.~Orvieto, L.~Gresele, B.~Sch{\"o}lkopf, Learning explanations that are hard to vary, ICLR (2021).

\bibitem{shi2021gradient}
Y.~Shi, J.~Seely, P.~H. Torr, N.~Siddharth, A.~Hannun, N.~Usunier, G.~Synnaeve, Gradient matching for domain generalization, ICLR (2022).

\bibitem{lakshminarayanan2017simple}
B.~Lakshminarayanan, A.~Pritzel, C.~Blundell, Simple and scalable predictive uncertainty estimation using deep ensembles, NeurIPS 30 (2017).

\bibitem{izmailov2018averaging}
P.~Izmailov, D.~Podoprikhin, T.~Garipov, D.~Vetrov, A.~G. Wilson, Averaging weights leads to wider optima and better generalization, UAI (2018).

\bibitem{arpit2022ensemble}
D.~Arpit, H.~Wang, Y.~Zhou, C.~Xiong, Ensemble of averages: Improving model selection and boosting performance in domain generalization, NeurIPS (2022).

\bibitem{wortsman2022model}
M.~Wortsman, G.~Ilharco, S.~Y. Gadre, R.~Roelofs, R.~Gontijo-Lopes, A.~S. Morcos, H.~Namkoong, A.~Farhadi, Y.~Carmon, S.~Kornblith, L.~Schmidt, Model soups: averaging weights of multiple fine-tuned models improves accuracy without increasing inference time (2022).

\bibitem{rame2022diverse}
A.~Rame, M.~Kirchmeyer, T.~Rahier, A.~Rakotomamonjy, patrick gallinari, M.~Cord, Diverse weight averaging for out-of-distribution generalization, NeurIPS (2022).

\bibitem{bommasani2021opportunities}
R.~Bommasani, D.~A. Hudson, E.~Adeli, R.~Altman, S.~Arora, S.~von Arx, M.~S. Bernstein, J.~Bohg, A.~Bosselut, E.~Brunskill, et~al., On the opportunities and risks of foundation models, arXiv preprint arXiv:2108.07258 (2021).

\bibitem{koh21a}
P.~W. Koh, S.~Sagawa, H.~Marklund, S.~M. Xie, M.~Zhang, A.~Balsubramani, W.~Hu, M.~Yasunaga, R.~L. Phillips, I.~Gao, T.~Lee, E.~David, I.~Stavness, W.~Guo, B.~Earnshaw, I.~Haque, S.~M. Beery, J.~Leskovec, A.~Kundaje, E.~Pierson, S.~Levine, C.~Finn, P.~Liang, Wilds: A benchmark of in-the-wild distribution shifts, ICML (2021) 5637--5664.

\bibitem{Zhang2023}
Y.~Ji, L.~Zhang, J.~Wu, B.~Wu, L.~Li, L.-K. Huang, T.~Xu, Y.~Rong, J.~Ren, D.~Xue, et~al., Drugood: Out-of-distribution dataset curator and benchmark for ai-aided drug discovery--a focus on affinity prediction problems with noise annotations, AAAI 37 (2023).

\bibitem{Tossou2023}
P.~Tossou, C.~Wognum, M.~Craig, H.~Mary, E.~Noutahi, Real-world molecular out-of-distribution: Specification and investigation, Journal of Chemical Information and Modeling (2024).

\bibitem{shepard2017developments}
H.~M. Shepard, G.~L. Phillips, C.~D. Thanos, M.~Feldmann, Developments in therapy with monoclonal antibodies and related proteins, Clinical medicine 17 (2017).

\bibitem{minot2024meta}
M.~Minot, S.~T. Reddy, Meta learning addresses noisy and under-labeled data in machine learning-guided antibody engineering, Cell Systems 15~(1) (2024).

\bibitem{hummer2023investigating}
A.~M. Hummer, C.~Schneider, L.~Chinery, C.~M. Deane, Investigating the volume and diversity of data needed for generalizable antibody-antigen $\delta$$\delta$g prediction, bioRxiv (2023) 2023--05.

\bibitem{olsen2022observed}
T.~H. Olsen, F.~Boyles, C.~M. Deane, Observed antibody space: A diverse database of cleaned, annotated, and translated unpaired and paired antibody sequences, Protein Science 31 (2022).

\bibitem{dunbar2014sabdab}
J.~Dunbar, K.~Krawczyk, J.~Leem, T.~Baker, A.~Fuchs, G.~Georges, J.~Shi, C.~M. Deane, Sabdab: the structural antibody database, Nucleic acids research 42 (2014).

\bibitem{raybould2020thera}
M.~I. Raybould, C.~Marks, A.~P. Lewis, J.~Shi, A.~Bujotzek, B.~Taddese, C.~M. Deane, Thera-sabdab: the therapeutic structural antibody database, Nucleic acids research 48 (2020).

\bibitem{schneider2022sabdab}
C.~Schneider, M.~I. Raybould, C.~M. Deane, Sabdab in the age of biotherapeutics: updates including sabdab-nano, the nanobody structure tracker, Nucleic acids research 50 (2022).

\bibitem{Hummer2023}
A.~M. Hummer, C.~Schneider, L.~Chinery, C.~M. Deane, Investigating the volume and diversity of data needed for generalizable antibody-antigen ddg prediction, bioRxiv (2023).

\bibitem{saremi2019neural}
S.~Saremi, A.~Hyvarinen, Neural empirical bayes, JLMR 20 (2019).

\bibitem{barlow2018flex}
K.~A. Barlow, S.~O~Conchuir, S.~Thompson, P.~Suresh, J.~E. Lucas, M.~Heinonen, T.~Kortemme, Flex ddg: Rosetta ensemble-based estimation of changes in protein--protein binding affinity upon mutation, The Journal of Physical Chemistry B 122 (2018).

\bibitem{chaudhury2010pyrosetta}
S.~Chaudhury, S.~Lyskov, J.~J. Gray, Pyrosetta: a script-based interface for implementing molecular modeling algorithms using rosetta, Bioinformatics 26 (2010).

\bibitem{sirin2016ab}
S.~Sirin, J.~R. Apgar, E.~M. Bennett, A.~E. Keating, Ab-bind: antibody binding mutational database for computational affinity predictions, Protein Science 25 (2016).

\bibitem{mahajan2022hallucinating}
S.~P. Mahajan, J.~A. Ruffolo, R.~Frick, J.~J. Gray, Hallucinating structure-conditioned antibody libraries for target-specific binders, Frontiers in immunology 13 (2022).

\bibitem{faure2022mapping}
A.~J. Faure, J.~Domingo, J.~M. Schmiedel, C.~Hidalgo-Carcedo, G.~Diss, B.~Lehner, Mapping the energetic and allosteric landscapes of protein binding domains, Nature 604 (2022).

\bibitem{honegger2001yet}
A.~Honegger, A.~Plu{\`E}ckthun, Yet another numbering scheme for immunoglobulin variable domains: an automatic modeling and analysis tool, Journal of molecular biology 309 (2001).

\bibitem{he2016deep}
K.~He, X.~Zhang, S.~Ren, J.~Sun, Deep residual learning for image recognition, CVPR (2016).

\bibitem{lin2023evolutionary}
Z.~Lin, H.~Akin, R.~Rao, B.~Hie, Z.~Zhu, W.~Lu, N.~Smetanin, R.~Verkuil, O.~Kabeli, Y.~Shmueli, et~al., Evolutionary-scale prediction of atomic-level protein structure with a language model, Science 379 (2023).

\bibitem{zhang2022protein}
Z.~Zhang, M.~Xu, A.~Jamasb, V.~Chenthamarakshan, A.~Lozano, P.~Das, J.~Tang, Protein representation learning by geometric structure pretraining, ICLR (2023).

\bibitem{jumper2021highly}
J.~Jumper, R.~Evans, A.~Pritzel, T.~Green, M.~Figurnov, O.~Ronneberger, K.~Tunyasuvunakool, R.~Bates, A.~{\v{Z}}{\'\i}dek, A.~Potapenko, et~al., Highly accurate protein structure prediction with alphafold, Nature 596~(7873) (2021) 583--589.

\bibitem{cha2021swad}
J.~Cha, S.~Chun, K.~Lee, H.-C. Cho, S.~Park, Y.~Lee, S.~Park, {SWAD}: Domain generalization by seeking flat minima, NeurIPS (2021).

\bibitem{baehrens2010explain}
D.~Baehrens, T.~Schroeter, S.~Harmeling, M.~Kawanabe, K.~Hansen, K.-R. M{\"u}ller, How to explain individual classification decisions, JMLR 11 (2010).

\bibitem{simonyan2013deep}
K.~Simonyan, A.~Vedaldi, A.~Zisserman, Deep inside convolutional networks: Visualising image classification models and saliency maps, CVPR (2014).

\bibitem{morea1998conformations}
V.~Morea, A.~Tramontano, M.~Rustici, C.~Chothia, A.~M. Lesk, Conformations of the third hypervariable region in the vh domain of immunoglobulins, Journal of molecular biology 275 (1998).

\bibitem{teplyakov2014canonical}
A.~Teplyakov, G.~L. Gilliland, Canonical structures of short cdr-l3 in antibodies, Proteins: Structure, Function, and Bioinformatics 82 (2014).

\bibitem{melnyk2023reprogramming}
I.~Melnyk, V.~Chenthamarakshan, P.-Y. Chen, P.~Das, A.~Dhurandhar, I.~Padhi, D.~Das, Reprogramming pretrained language models for antibody sequence infilling, ICML (2023).

\bibitem{prihoda2022biophi}
D.~Prihoda, J.~Maamary, A.~Waight, V.~Juan, L.~Fayadat-Dilman, D.~Svozil, D.~A. Bitton, Biophi: A platform for antibody design, humanization, and humanness evaluation based on natural antibody repertoires and deep learning, MAbs 14 (2022).

\bibitem{giudicelli2005imgt}
V.~Giudicelli, D.~Chaume, M.-P. Lefranc, Imgt/gene-db: a comprehensive database for human and mouse immunoglobulin and t cell receptor genes, Nucleic acids research 33 (2005).

\bibitem{buchfink2021sensitive}
B.~Buchfink, K.~Reuter, H.-G. Drost, Sensitive protein alignments at tree-of-life scale using diamond, Nature methods 18 (2021).

\bibitem{jorgensen2008perspective}
W.~L. Jorgensen, L.~L. Thomas, Perspective on free-energy perturbation calculations for chemical equilibria, Journal of chemical theory and computation 4 (2008).

\bibitem{clark2017free}
A.~J. Clark, T.~Gindin, B.~Zhang, L.~Wang, R.~Abel, C.~S. Murret, F.~Xu, A.~Bao, N.~J. Lu, T.~Zhou, et~al., Free energy perturbation calculation of relative binding free energy between broadly neutralizing antibodies and the gp120 glycoprotein of hiv-1, Journal of molecular biology 429 (2017).

\bibitem{clark2019relative}
A.~J. Clark, C.~Negron, K.~Hauser, M.~Sun, L.~Wang, R.~Abel, R.~A. Friesner, Relative binding affinity prediction of charge-changing sequence mutations with fep in protein--protein interfaces, Journal of molecular biology 431 (2019).

\bibitem{zhu2022large}
F.~Zhu, F.~A. Bourguet, W.~F. Bennett, E.~Y. Lau, K.~T. Arrildt, B.~W. Segelke, A.~T. Zemla, T.~A. Desautels, D.~M. Faissol, Large-scale application of free energy perturbation calculations for antibody design, Nature Scientific Reports 12 (2022).

\bibitem{mason2021optimization}
D.~M. Mason, S.~Friedensohn, C.~R. Weber, C.~Jordi, B.~Wagner, S.~M. Meng, R.~A. Ehling, L.~Bonati, J.~Dahinden, P.~Gainza, et~al., Optimization of therapeutic antibodies by predicting antigen specificity from antibody sequence via deep learning, Nature Biomedical Engineering 5 (2021).

\bibitem{pushkarna2022data}
M.~Pushkarna, A.~Zaldivar, O.~Kjartansson, Data cards: Purposeful and transparent dataset documentation for responsible ai, in: Conference on Fairness, Accountability, and Transparency, 2022.

\end{thebibliography}
\bibliographystyle{elsarticle-num}


\newpage
\appendix

\newpage

\tableofcontents

\section{Additional results from experiments}

\subsection{Finetuning ESM - detailed implementation}
\textbf{Finetuned ESM2:} We finetune the 8M-parameter ESM2 \cite{lin2023evolutionary}, a protein language model pretrained on experimental and high-quality predicted structures of general proteins. 
We used a single ESM2 model for the two antibody chains as well as the antigen. One potential challenge with fine-tuning a single ESM2 model on three protein chains is that they are OOD for ESM2, which was pretrained on single chains. To address this, we follow the tricks used in ESMFold \citep{lin2023evolutionary}: (1) adding a 25-residue poly-glycine linker between the heavy and light antibody chains and between the antibody and antigen and (2) implementing a jump in residue index into the positional embeddings at the start of each new chain. The tricks significantly boosted the classifier performance, signaling that the structure information in the ESM embeddings was important. 

\subsection{Model selection: test-domain validation set}
\label{sec:app_results}
\begin{table}[htbp!]
\caption{Model selection: test-domain validation set, corresponding to the training domain validation set in the main text. \label{tab:test_res}}
\begin{center}
\adjustbox{max width=\textwidth}{%
\begin{tabular}{lccccc}
\toprule
\textbf{Algorithm}   & \textbf{Env 0}     & \textbf{Env 1}     & \textbf{Env 2}     & \textbf{Env 3}     & \textbf{Avg}         \\
\midrule
random                  & 50.5 $\pm$ 0.0       & 52.4 $\pm$ 5.2       & 38.8 $\pm$ 0.0       & 49.1 $\pm$ 1.1       & 47.7                 \\
\midrule
 & &   \textbf{ESM} &&\\
\midrule
ERM                  & 60.5 $\pm$ 0.9       & 67.6 $\pm$ 0.6       & 69.3 $\pm$ 0.5       & 66.1 $\pm$ 0.5       & 65.9                 \\
ERM-ENS                  &  62.1     & 69.1       &  73.2      & 69.2       & 68.4                \\
SMA              & 62.9 $\pm$ 0.8       & 68.3 $\pm$ 0.4       & 70.7 $\pm$ 0.5       & 66.7 $\pm$ 0.4       & 67.1                 \\
SMA-ENS             &  63.8       &  70.0      &  72.8      & 68.3       & 68.7                 \\
IRM                  & 59.2 $\pm$ 0.5       & 65.5 $\pm$ 1.8       & 68.6 $\pm$ 0.3       & 63.6 $\pm$ 1.1       & 64.2                 \\
IRM-ENS                  & 62.3       & 68.9       & 70.4       & 66.9       & 67.1                 \\
CORAL                & 60.7 $\pm$ 0.1       & 67.5 $\pm$ 0.1       & 68.4 $\pm$ 0.2       & 66.8 $\pm$ 0.3       & 65.8                 \\
CORAL-ENS                  & 62.6       & 70.5       & 71.9       & 67.7       & 68.2                  \\
VREx                 &  60.4 $\pm$ 1.1       & 65.4 $\pm$ 0.9       & 69.6 $\pm$ 0.5       & 65.9 $\pm$ 0.2       & 65.3                    \\
VREx-ENS                  & 62.4     & 68.3       & 73.0     & 68.1    &  68.0                \\
Fish                 & 61.0 $\pm$ 1.6       & 65.9 $\pm$ 0.5       & 70.6 $\pm$ 1.3       & 66.3 $\pm$ 0.3       & 66.2                    \\
\midrule
 & &  \textbf{SeqCNN} &&\\
 \midrule

ERM                  & 61.4 $\pm$ 1.4       & 64.4 $\pm$ 0.1       & 66.5 $\pm$ 0.6       & 63.8 $\pm$ 0.5       & 64.0                 \\
ERM-ENS &     62.6 &              66.2           &   68.0  &          66.5&  65.8\\
SMA              & 57.0 $\pm$ 0.4       & 65.9 $\pm$ 0.3       & 65.8 $\pm$ 0.2       & 64.5 $\pm$ 0.4       & 63.3                                \\
SMA-ENS            & 61.2     & 67.0       & 67.1      & 66.2    & 65.4                                \\
IRM                  & 61.7 $\pm$ 1.4       & 65.8 $\pm$ 0.5       & 68.6 $\pm$ 1.6       & 63.7 $\pm$ 0.8       & 64.9                 \\
IRM-ENS                  & 62.70       & 64.34       & 63.51        &   71.56  & 65.5                \\
CORAL                & 58.3 $\pm$ 1.5       & 65.3 $\pm$ 0.5       & 65.9 $\pm$ 0.9       & 63.2 $\pm$ 0.8       & 63.2                 \\
CORAL-ENS                & 62.0        & 67.4       & 68.4      & 65.4      & 65.8                 \\
VREx                 & 58.4 $\pm$ 1.4       & 65.3 $\pm$ 0.3       & 65.1 $\pm$ 0.3       & 63.8 $\pm$ 0.4       & 63.2                 \\
VREx-ENS                 & 60.9    & 66.8       & 67.83       & 66.73    & 65.6                 \\
Fish                 & 59.9 $\pm$ 2.6       & 67.0 $\pm$ 0.1       & 69.4 $\pm$ 0.4       & 65.3 $\pm$ 0.4       & 65.4                                 \\
\midrule
 & &  \textbf{GearNet} & &\\
\midrule
ERM                 &  62.7 $\pm$ 1.6       & 79.7 $\pm$ 0.4       & 85.9 $\pm$ 1.2       & 76.0 $\pm$ 0.5       & 76.1  \\
ERM-ENS                 & \textbf{64.3}       & \textbf{81.7}      & \textbf{88.6}       & \textbf{78.1}      & \textbf{78.2}                 \\
VREx                 & 58.0 $\pm$ 0.1       & 73.8 $\pm$ 1.0       & 82.3 $\pm$ 1.5       & 72.5 $\pm$ 0.3       & 71.6                 \\
VREx-ENS                 &  62.2      & 77.3      & 86.9       & 74.6      & 75.3                 \\
\bottomrule
\end{tabular}}
\end{center}
\end{table}

\subsection{IRM v1 penalty} \label{app:irm_v1_penalty}

\begin{figure}[t]
\centering
\includegraphics[width=0.5\textwidth]{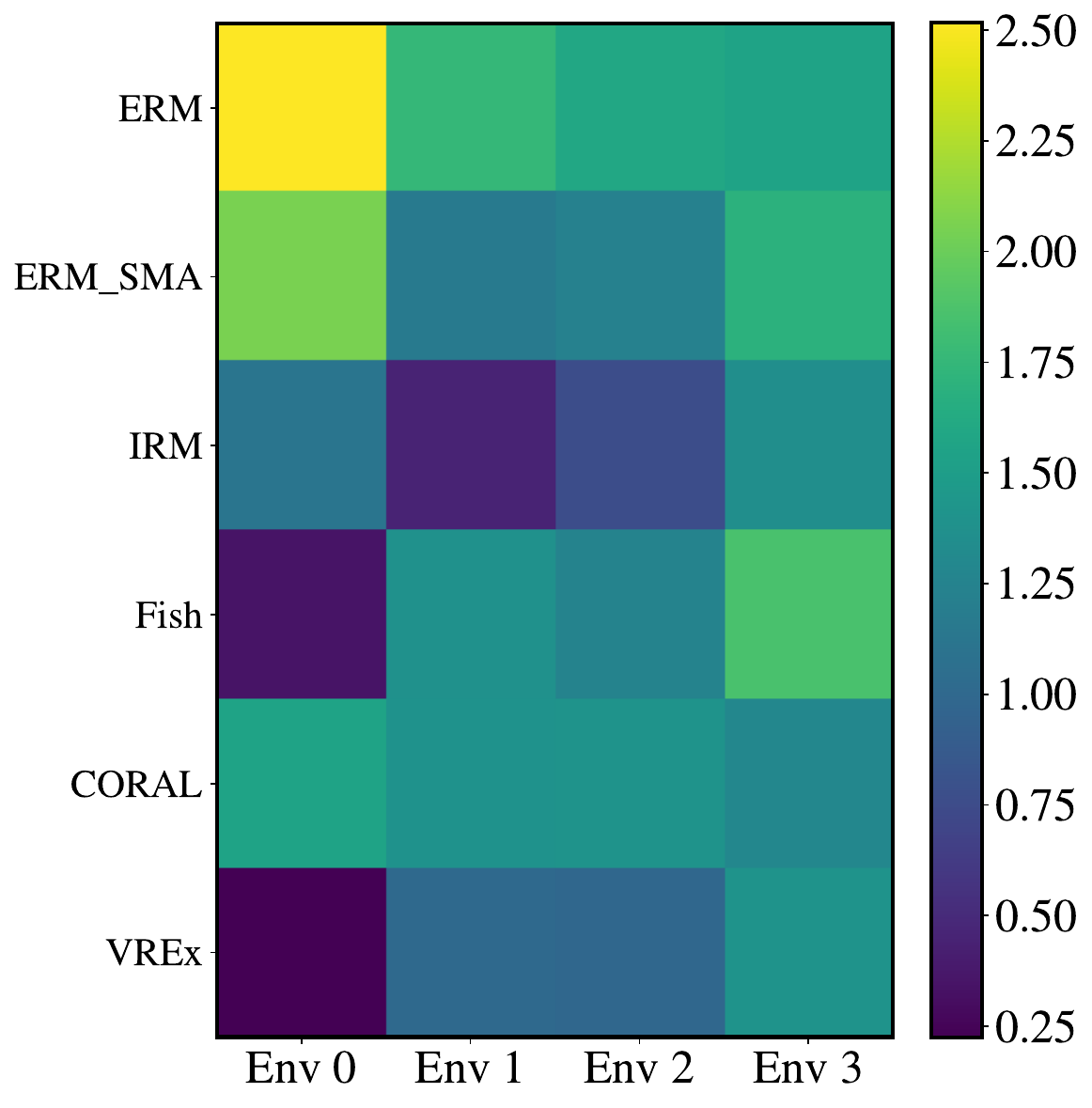}
\caption{ERM has the highest IRM v1 penalty across all environments. The highest penalty value for ERM in Env 0 is consistent with the accuracy for Env 0 being the lowest (see \autoref{tab:train_res}, \autoref{tab:test_res}). \label{fig:irm_v1_penalty}}
\end{figure}

The IRM v1 is a relaxation of IRM originally proposed by \cite{arjovsky2019invariant}, where the classifier $w$ is fixed to a scalar 1. The loss function includes a penalty term $||\nabla_{w|w=1} R_e(w \cdot \Phi)||^2$. Theorem 4 of \cite{arjovsky2019invariant} is used to justify the use of this term as the invariance penalty for all differentiable loss functions, such as the cross-entropy loss function used in this paper. \autoref{fig:irm_v1_penalty} plots this invariance penalty for various algorithms across the four environments. IRM carries the least IRM v1 penalty overall, as expected. Other invariance-based algorithms also carry low penalty values, while ERM and ERM-SMA have high penalty values, particularly in Env 0. The high relative penalty value in Env 0 is consistent with the accuracy for Env 0 being the lowest (see \autoref{tab:train_res}, \autoref{tab:test_res}).

\subsection{Platform specifications}
All experiments have been executed on NVIDIA A100 Tensor Core GPU.

\subsection{Generalizing to new antigen targets}
\label{sec:app_her2}

One challenging yet practical scenario is being able to predict a property of interest for a new antigen that has not been seen during training. We were thus motivated to evaluate the DG algorithms to an unseen target, HER2. The data in this new domain, consisting only of HER2 designs, was obtained from Graphinity \citep{Hummer2023}, which significantly differs from the WJS generative model in the distribution of antibody sequences. Graphinity it is a brute-force method that starting from a seed sequence, applies all possible point mutations at each of the positions in the CDR3 loop of an antibody. Which means that this method produces antibodies with edit distance of only 1 (one amino acid different from the starting seed). This difference in sequence distance amounts to a change in covariate and sub-population shift, compared to the other environments with WJS designs. Additionally, since these designs were scored using a different $\Delta \Delta G$ model, FoldX, this environment also includes label shift. With so much distribution shift compared to the rest of the data, it is expected that models trained on the other targets (HIV and SARS-Cov-2) will not generalize to HER2 designs. However, we wanted to investigate if the DG algorithms will have some advantage over vanilla ERM.

The table below summarizes the results where environments consist of the data curated as described in section \autoref{sec:domain_bed}.
The results unfortunately are not in favour of any of the DG algorithms. There is an advantage of the SeqCNN framework achieving higher accuracy on environment 4, however, those number are still around 50\% and hence we can not consider them useful since we could not use such model in practice.

We are further investigating if including a new target but without label shift (i.e. using the pyRosetta scores) may deliver better results. As previously mentioned, having a model that can reliably predict binding or other molecular properties while being antigen agnostic is of crucial importance in accelerating drug discovery and design.

\begin{table}[H]

\begin{center}
\caption{Model selection: train-domain validation set when HER2 is included.}
    \adjustbox{max width=0.8\textwidth}{%
    \begin{tabular}{lcccccc}
    \toprule
    \textbf{Algorithm}   & \textbf{env-0}     & \textbf{env-1}     & \textbf{env-2}     & \textbf{env-3}     & \textbf{env-4}     & \textbf{Avg}         \\

    \midrule
     & &  & \textbf{SeqCNN} &\\\midrule
ERM & 63.2 $\pm$ 1.0       & 66.2 $\pm$ 0.9       & 66.2 $\pm$ 0.8       & 64.9 $\pm$ 0.1         & 46.6 +/- 3.5       &   61.42               \\
ERM-ENS   &     62.3      &          66.1          &    69.7             &  65.9  &   51.93    &       63.18       \\
SMA              & 61.8 $\pm$ 0.9       & 66.5 $\pm$ 0.3       & 66.1 $\pm$ 0.2       & 64.9 $\pm$ 0.3      &  51.9 +/- 1.5          &   62.24            \\
SMA -ENS             & 58.6    &   66.9     & 68.8      & 66.1      & \textbf{55.71}   &  \textbf{63.22}    \\
IRM                  &  60.0 $\pm$ 0.9       & 64.4 $\pm$ 0.2       & 69.6 $\pm$ 0.9       & 63.5 $\pm$ 0.6       & 39.1 +/- 4.4         &            59.32   \\
IRM-ENS                  &  62.4      &  66.5      & 73.2 &65.1    & 43.06    &   59.32    \\
VREx                 & 60.1 $\pm$ 1.6       & 65.7 $\pm$ 1.0       & 66.3 $\pm$ 0.6  &64.9 $\pm$ 0.4  & 49.7 +/- 1.8        &   61.34           \\
VREx -ENS                & 61.5      & 66.9       & 68.2    &  66.1& 48.85   &      62.31   \\
Fish                 & 58.2 $\pm$ 1.3       & 66.0 $\pm$ 0.2       & 68.2 $\pm$ 0.6 &65.8 $\pm$ 0.3       & 51.5 $\pm$ 2.2       & 61.94                 \\
    \bottomrule
    \end{tabular}}
\end{center}
\end{table}

\begin{table}[H]
\caption{Model selection: test-domain validation set (oracle) when HER2 is included.}
\begin{center}
\adjustbox{max width=0.8\textwidth}{%
\begin{tabular}{lcccccc}
\toprule
\textbf{Algorithm}   & \textbf{env-0}     & \textbf{env-1}     & \textbf{env-2}     & \textbf{env-3}     & \textbf{env-4}     & \textbf{Avg}         \\
    \midrule
     & &  & \textbf{SeqCNN} &\\  \midrule
ERM                  & 61.4 $\pm$ 1.4       & 64.4 $\pm$ 0.1       & 66.5 $\pm$ 0.6       & 63.8 $\pm$ 0.5            & 54.5 +/- 1.1        &           62.12     \\
ERM-ENS          &       62.6 &              66.2           &   68.0  &     65.9&     38.41 &        60.22      \\
SMA              & 57.0 $\pm$ 0.4       & 65.9 $\pm$ 0.3       & 65.8 $\pm$ 0.2       & 64.5 $\pm$ 0.4         &  50.7 +/- 1.7        & 60.78         \\
SMA-ENS              &   61.2     & 67.0       & 67.1      & 66.2    &45.07 &           61.34    \\
IRM                  & 61.7 $\pm$ 1.4       & 65.8 $\pm$ 0.5       & 68.6 $\pm$ 1.6       & 63.7 $\pm$ 0.8&58.9 +/- 2.3      &63.74            \\
IRM-ENS                  & 62.70       & 64.34       & 63.51        &   71.56   & 59.53   &     64.32        \\
VREx                 & 58.4 $\pm$ 1.4       & 65.3 $\pm$ 0.3       & 65.1 $\pm$ 0.3       & 63.8 $\pm$ 0.4            &53.6 +/- 0.1       & 61.24               \\
VREx-ENS                &      60.9    & 66.8       & 67.83       & 66.73  &   40.47&      60.54        \\
Fish                 & 59.9 $\pm$ 2.6       & 67.0 $\pm$ 0.1       & 69.4 $\pm$ 0.4       & 65.3 $\pm$ 0.4      &52.0 +/- 1.2         &  62.72   \\
\bottomrule
\end{tabular}}
\end{center}
\end{table}

\section{Dataset properties}
\label{sec:app_ab_valid}
In this section we evaluate the validity of the antibodies in our synthetic library. We do so, to ensure quality and reliability of the proposed benchmark. We evaluate the following properties:
\begin{enumerate}
    \item \emph{naturalness} - measures the similarity of the antibody sequences in Antibody Domainbed to antibodies in OAS  \citep{olsen2022observed}, the largest publicly available database of natural antibodies. These scores come from the perplexity of a language model trained on the OAS database \citep{melnyk2023reprogramming}. Higher is better.
    \item \emph{OASis -Percentile/Identity/Germline Content} we include three scores computed with the recent publicly available humanness evaluation platform BioPhi \citep{prihoda2022biophi}. We used the default setup in relaxed mode. Briefly, OASis identity score for an input sequence is calculated as the fraction of peptides with prevalence in therapeutic antibodies curated from OAS \citep{olsen2022observed}. OASis Percentile converts the identity score to 0-100 range based on therapeutic antibodies such that 0\% OASis percentile score corresponds to the least human and the 100\% OASis percentile score corresponds to the most human antibody in the clinic. Germline content is yet another humanness score, which represents the percent sequence identity with a concatenation of the nearest human V and J gene  (percent sequence identity with nearest human germline) with regards to IMGT Gene-DB \citep{giudicelli2005imgt}.
    \item \emph{bio-informatics properties} hydrophobicity - a measure of the degree of affinity between water and a side chain of an amino acid; pi charge - the pH at which the antibody has no net electrical charge, this value depends on the charged amino acids the antibody contains; and aromaticity - binding of the two complementary surfaces is mostly driven by aromatic residues. For these three properties we use the corresponding  bioPython implementations, and we compare the range of values to in-vitro functional antibodies (env 5).
    \item \emph{diamond} we use this score to explore closeness of the proposed designs to the OAS database, by fast sequence alignment inspired by \citep{buchfink2021sensitive}. Higher scores are preffered. 
    
\end{enumerate}

\begin{figure}[ht!]
    \centering
    \includegraphics[width=0.95\textwidth]{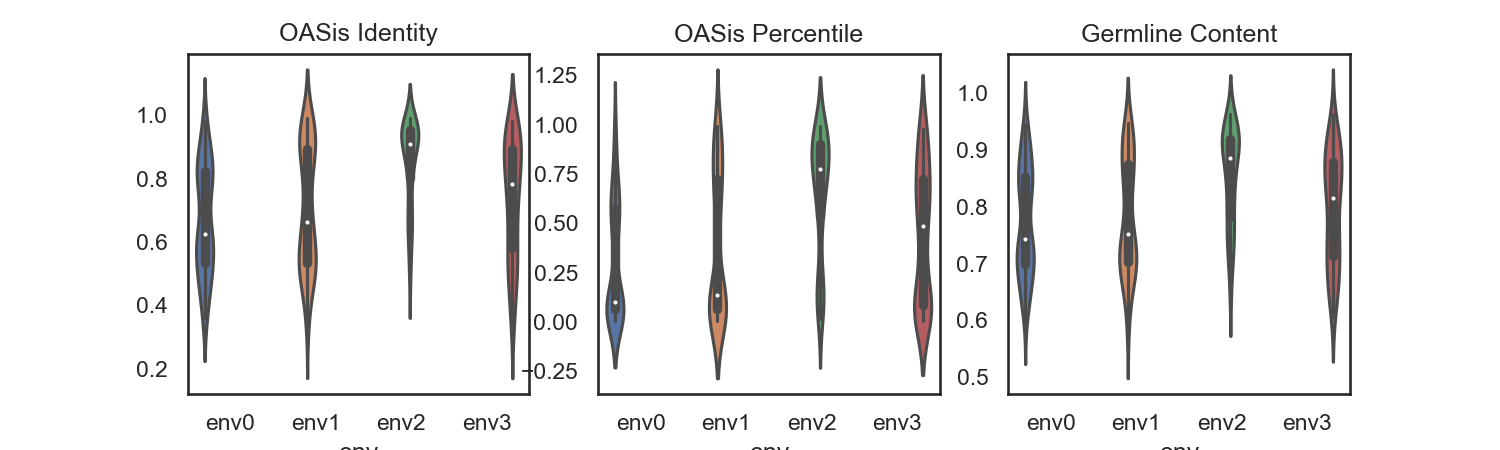}
    \caption{OASis properties for Antibody domainbed computed using BioPhi \citep{prihoda2022biophi}. Note that env 5 was omitted due to proprietary reasons.}
    \label{fig:oasis}
    
\end{figure}

\begin{figure}
    \centering
    \includegraphics[width=0.85\textwidth]{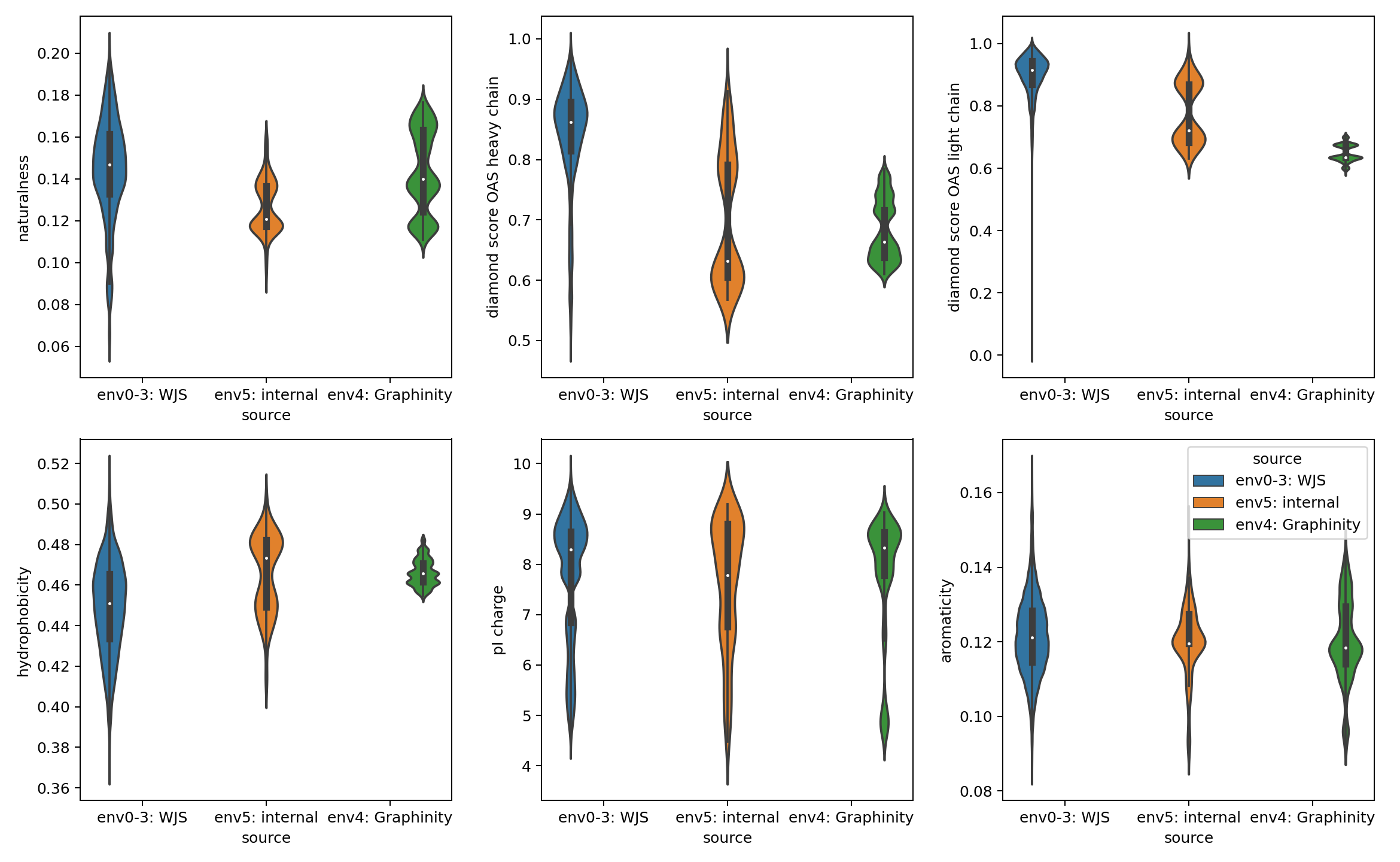}
    \caption{Various properties for evaluating quality of synthetic antibodies. Please see text for details.}
    \label{fig:properties}
    
\end{figure}

Our results are presented in \autoref{fig:properties} and \autoref{fig:oasis}. 
In \autoref{fig:properties}, first row, naturalness and diamond score, we confirm that WJS generated antibodies (Env 0-3) have properties close to observed antibodies, and even more so, they achieve better scores than the single point mutations in Graphinity (env4) and the human-expert designs from internal in-vitro experiments (env5). Next, in the second row of 
\autoref{fig:properties}, we notice that the ranges of values for hydrophobicity, charge and aromaticity mostly overlap between WJS antibodies and in-vitro functional measurements (env5). These results reconfirm what was already included in the original WJS publication \cite{frey2023learning}. In \autoref{fig:oasis} we investigate the humanness of the proposed antibodies. Note that since this platform requires uploading of sequences, we were not in position to score our internal experimental sequences. Hence, we provide the results only for Env 0-3. These results confirm that the WJS antibodies provide a well-balanced mix of antibodies both close and far to the therapeutic human reference dataset (cf. Figure 3 in \citep{prihoda2022biophi}), as expected within a drug design pipeline which leverages immunization campaigns (lead molecules from animal germlines) and ML generative models.

With this analysis in place, we are confident that our antibody benchmark is indeed representative of what we can expect in a real-world drug design framework.

\subsection{$\Delta \Delta G$ labels as proxy for affinity measurements} \label{app:ddg}
The change in Gibbs free energy, $\Delta G$, and the dissociation constant, $K_D$, can be shown to be theoretically equivalent up to a proportionality; we have $\Delta G = RT \ln K_D$, where $R$ is the gas constant, $1.98722 {\rm cal}/K\cdot {\rm mol}$, and $T$ is the absolute temperature \citep{jorgensen2008perspective}. Binding free energies have been applied to identify mutant antibodies with high affinity \citep{clark2017free, clark2019relative, zhu2022large}, supporting the use of $\Delta \Delta G$ as a synthetic proxy for $K_D$. 

That said, free energy is not exactly computable and we have used Rosetta and FoldX scores as weak approximations. The distribution of Rosetta-computed $\Delta \Delta G$ is still well-separated for binders and non-binders from fluorescence-activated cell sorting (FACS) and the separation signal is even stronger for binders from surface plasmon resonance (SPR) \citep{mason2021optimization,mahajan2022hallucinating}. (FACS is a higher-throughput but noisier method of identifying binders compared to SPR.)

In the case of environments 0 - 3, there is significant noise in the computed $\Delta \Delta G$ between -1 and 1 kcal/mol, because Rosetta and FoldX are less accurate at predicting $\Delta \Delta G$ for mutations with only a small effect on binding \citep{sirin2016ab,Hummer2023}. We therefore remove highly uncertain labels between -0.1 and 0.1 kcal/mol before attaching binary labels: label 1 if $\Delta\Delta G < -0.1$ (stabilizing) and  label 0 if $\Delta\Delta G > 0.1$ (destabilizing). We follow the following sign convention:
\begin{align} \label{eq:ddg_sign}
\Delta \Delta G  = \Delta G_\textrm{wild type} - \Delta G_{\rm mutant},
\end{align}
such that negative $\Delta \Delta G$ is stabilizing. Note that this represents a sign flip relative to the convention followed by \cite{Hummer2023}.

To validate the benefit of $\Delta \Delta G$ labels for predicting experimental binding measurements, we introduce environment 5, with details and description in \autoref{tab:ds}. 
This environment consists solely of antibodies targeting a variant of the HER2 antigen. Binding labels were obtained from internally conducted surface plasmon resonance (SPR) experiments. 

These variants overlap partially with antigen sequences in environment 4 (overlap between 30 - 70\%); environment 4 is a subset of Graphinity \citep{Hummer2023} and consists of synthetic single point designs aimed at HER2. The antibody sequences did not overlap with environment 4.

We report two baselines for evaluating the usefulness of training on  $\Delta \Delta G$ labels:
\begin{itemize}
    \item a random classifier.
    \item a binding affinity classifier, trained on ~5K pairs of antibodies and binding measurements for 4 internal targets (none of them HER2). This baseline mimicks the vanilla setup in drug discovery, where predictive models are confronted to a zero-shot setting where they are evaluated on new targets that differ from previous in-vitro experiments. This binding affinity classifier has accuracy of 0.64, precision 0.49 and recall 0.5.  
\end{itemize}

We then run Antibody Domain bed as previously, training on Env 0-4 but now evaluating on env 5. Our results are included in Table 6 and 7. We highlight two main results:
\begin{itemize}
    \item All algorithms, including ERM achieve better results than the two baselines (binding classifier and random). This reconfirms the benefit of $\Delta \Delta G$ labels. 
    \item CORAL, and CORAL-ENS achieve highest ac curacy, confirming the  benefit of leveraging the DG algorithms and our Antibody Domainbed benchmark.
\end{itemize}

\subsection{Sequence distance across environments in Antibody Domainbed}
\label{sec:seq_sim}

In what follows we compare the different environments in terms of sequence similarity. A common unit for comparison in the antibody design space is the edits, or sequence distance which is a discrete value representing number of positions with different amino-acids between two or more antibody (protein) sequences.

As we tried to separate the effect of the different generative models, by placing their corresponding designs into different environments, such split also amounts to gradually increasing the sequence distance to the seeds as the environment number progresses. From \autoref{fig:edist_per_model} we notice that highest sigma environment (Env 2, WJS $\sigma=2$) include the smaller sequence distance environments (Env 0 and 1, WJS $\sigma \in \lbrace 0.5, 1, 1.5 \rbrace $. 

Intuitively, smaller distances between sequences should amount to similar properties, however such intuition has never been confirmed fully as there is always a counter example where even a single point mutation may destroy some property of the antibody, depending on where its positioning in the sequence as well as its interaction with other atoms in the molecule or its' surroundings. We also notice this paradox in our results, the smallest distance environments usually being the most challenging one for all DG models, regardless of the backbone or the fact that such sequences have also been generated by the  generative models in the other environments. 

\begin{figure}[h]
    \centering
    \includegraphics[width=0.85\textwidth]{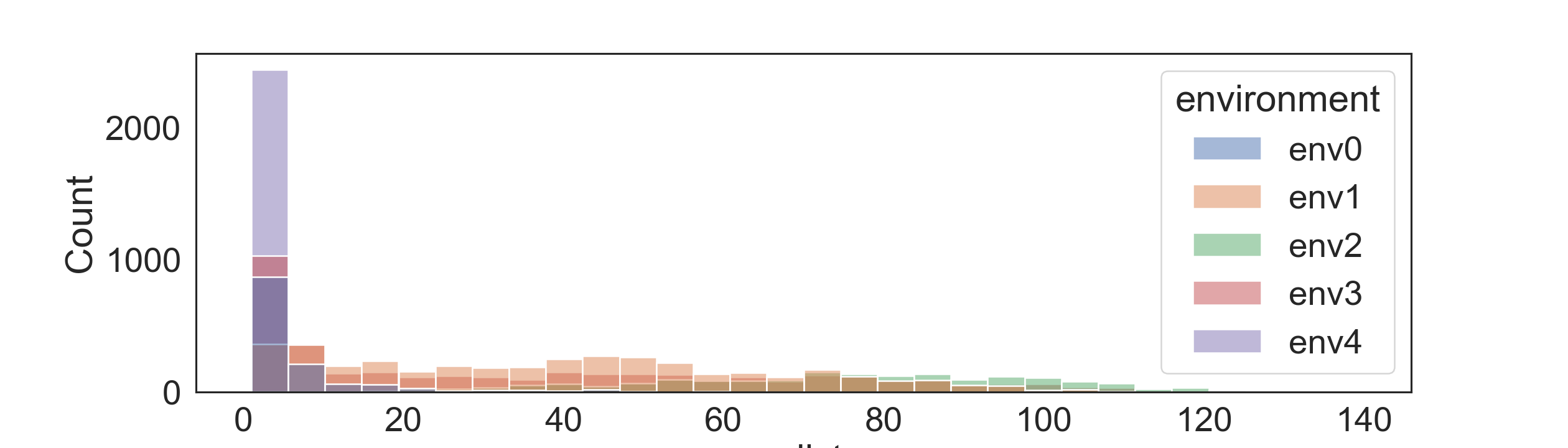}
    \caption{Sequence distances between antibody designs and their corresponding seeds. Colored by environment according to the split presented in \autoref{tab:ds}.}
    \label{fig:ab_struct}
\end{figure}

\begin{figure}[h]
    \centering
    \begin{subfigure}{\includegraphics[width=0.8\textwidth]{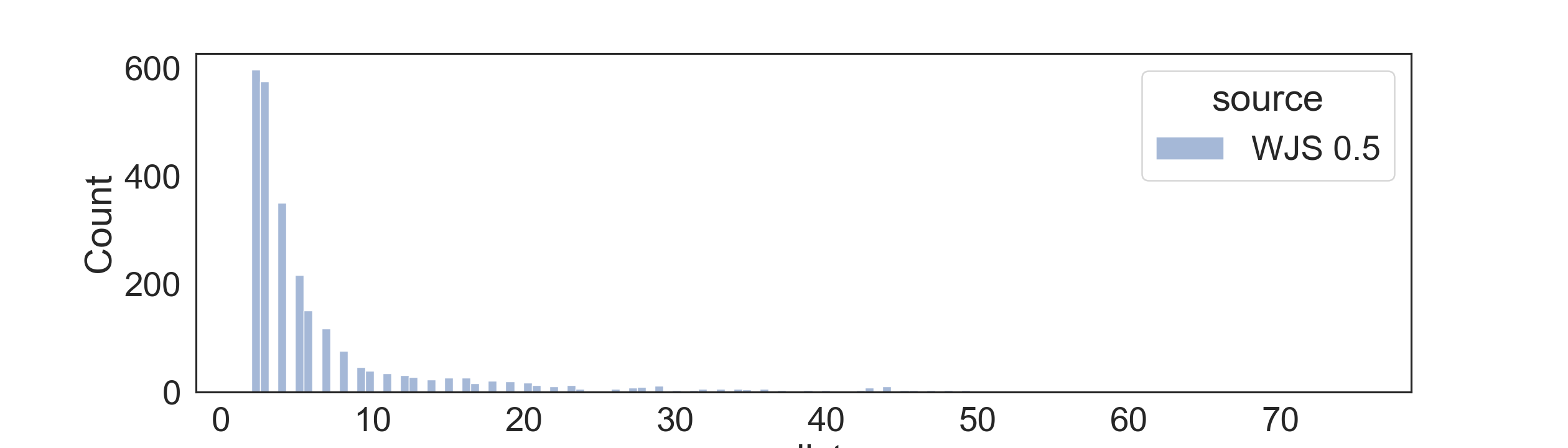}}
     \end{subfigure}
    \begin{subfigure}{\includegraphics[width=0.8\textwidth]{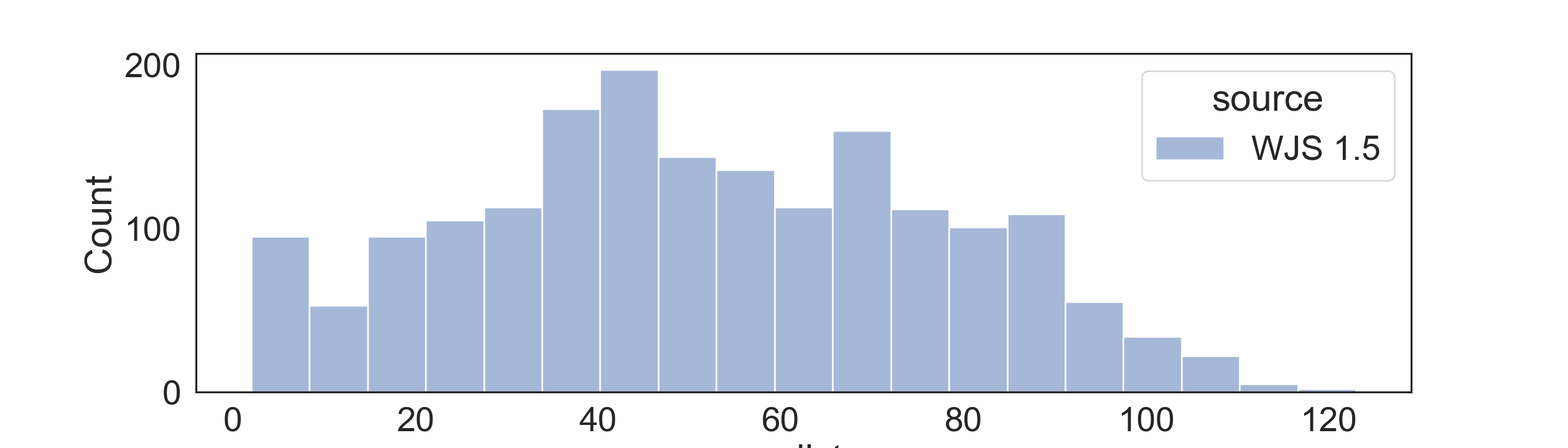}}
     \end{subfigure}
     \begin{subfigure}{\includegraphics[width=0.8\textwidth]{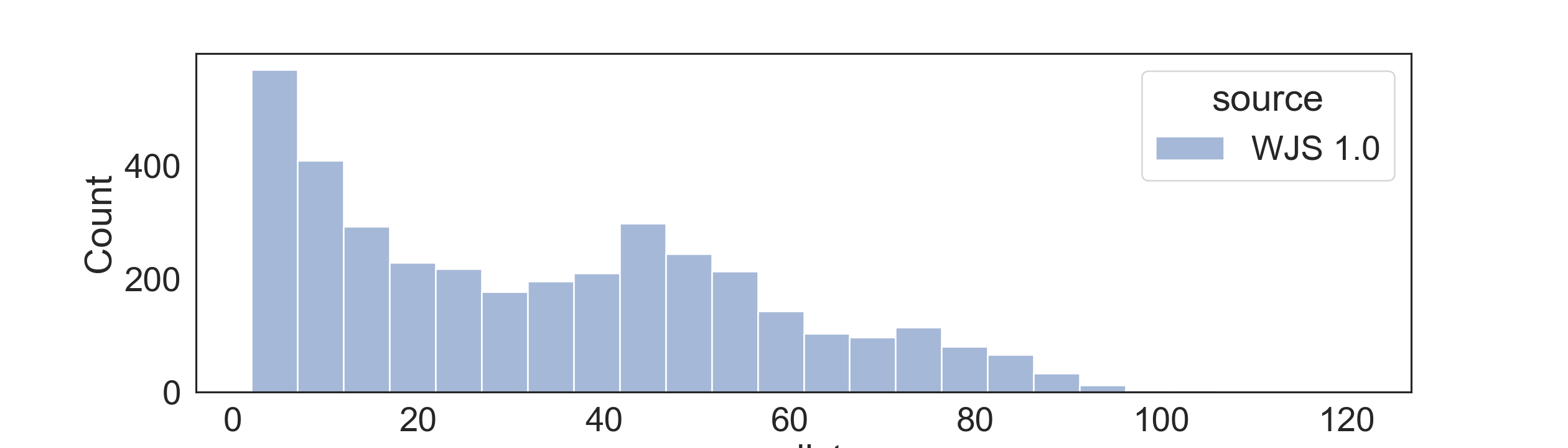}}
     \end{subfigure}
     \begin{subfigure}{\includegraphics[width=0.8\textwidth]{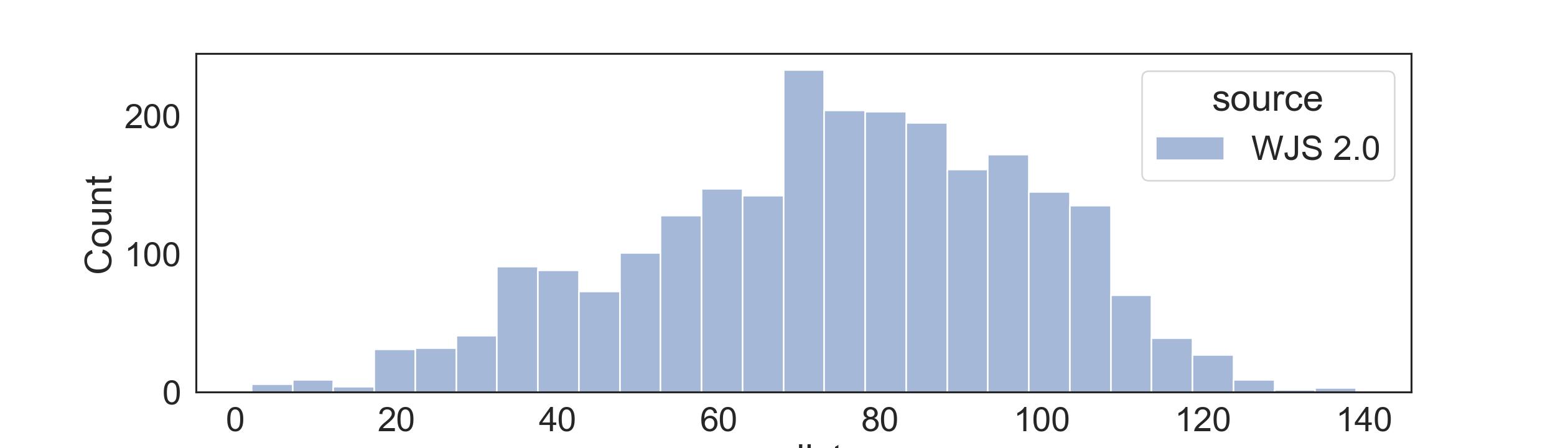}}
     \end{subfigure}
\caption{Sequence distances between antibody designs and their corresponding seeds for each WJS generative model. \label{fig:edist_per_model}}
 \end{figure}

\subsection{Model size robustness}
We additionally explored the necessity for larger models, by fine-tuning a 4x larger ESM model with 35M parameters. Due to memory issues, we had to reduce the batch size to 8 (while ESM 8M was run with batch size 12), and to compensate we increased the number of steps to 30 000 (ESM 8M had 15 000 steps). We repeated this experiment 3 times for 5 combinations of hyper parameters. Due to the computational intensity of this model and time constraints, we could only include three of the baseline in our current results. We don't notice any gain in the performance for Antibody Domainbed by increasing the model size.

\begin{table}[h!]
\caption{ESM 35M: Model selection with train and test-domain validation set.}
\label{}
\begin{center}
\adjustbox{max width=0.9\textwidth}{%
\begin{tabular}{lccccc}
\toprule
\textbf{Algorithm}   & \textbf{env0}     & \textbf{env1}     & \textbf{env2}     & \textbf{env3}     & \textbf{Average}         \\
\midrule 
& & \textbf{ESM 35M: train-domain}& & & \\
\midrule
ERM                  & 61.3 $\pm$ 1.7       & 66.5 $\pm$ 0.5       & 63.8 $\pm$ 3.7       & 64.5 $\pm$ 0.6       & 64.0                 \\
SMA              & 62.6 $\pm$ 1.6       & 66.8 $\pm$ 0.8       & 70.5 $\pm$ 0.0       & 63.1 $\pm$ 1.5       & 65.7                 \\
VREx                 & 61.9 $\pm$ 0.8       & 66.5 $\pm$ 0.0       & 65.9 $\pm$ 3.4       & 66.0 $\pm$ 0.2       & 65.1                 \\
VREx-ENS                  &  62.39  &   68.27     &     71.10   &      67.49  &   67.31              \\
\midrule
& & \textbf{ESM 35M: test-domain}& & & \\
\midrule
ERM                  & 60.6 $\pm$ 0.7       & 66.8 $\pm$ 0.7       & 60.0 $\pm$ 6.1       & 63.9 $\pm$ 0.4       & 62.8                 \\

SMA              & 62.8 $\pm$ 1.5       & 66.8 $\pm$ 1.2       & 69.2 $\pm$ 0.0       & 63.8 $\pm$ 2.0       & 65.6                 \\

VREx                 & 61.3 $\pm$ 0.8       & 66.7 $\pm$ 0.5       & 64.0 $\pm$ 4.4       & 65.8 $\pm$ 0.3       & 64.4                 \\
VREx-ENS                  &  60.78  &    68.34    &      71.27   &   67.14     &      66.9           \\

\bottomrule
\end{tabular}}
\end{center}
\end{table}

\section{Context - Therapeutic Antibodies}
\label{sec:ab_str}
Antibodies or immunoglobulins (Ig) maintain a common four -piece structure consisted of two identical heavy chains (HCs) and two identical light chains (LCs). The subunites are connected by disulfide bridges linking HCs and LCs together, forming the canonical ``Y" shape of antibodies.

The most important regions for antibody design are the variable domains (V) of both the heavy and light chains (VH and VL, respectively). These are the regions that interact with the antigens. These domains determine the specificity of an antibody (how likely it is to attach to other molecules in the body) through highly variable amino acid sequences. On the other hand, the constant domains (C) on heavy and light chains interact with effector proteins and molecules \autoref{fig:ab_struct}. 

On a more granular level, in the VH and VL domains, there are three complementarity-determining regions: CDR-H1, CDR-H2, and CDR-H3 for VH and CDR-L1, CDR-L2, CDR-L3 for VL. These subregions are highly variable in their amino acid sequences, and they form distinct loops creating a surface complementary to distinct antigens. CDR-H3 is known to be the main contributor to antigen recognition due to its sequence diversity, length, and favourable location. Since CDR-H3 loop has an impact on the loop conformations and antigen binding at the other CDRs, it is the main driver of specificity and binding affinity. In-between the CDRs, we see the framework regions (FR). Frameworks entail less variability and provide structural stability for the whole domain. The FR regions induce a $\beta$ sheet structure, at which the CDR loops are located at the outer edge, forming an antigen-binding site. 

\begin{figure}[htbp!]
    \centering
    \includegraphics[width=0.8\textwidth]{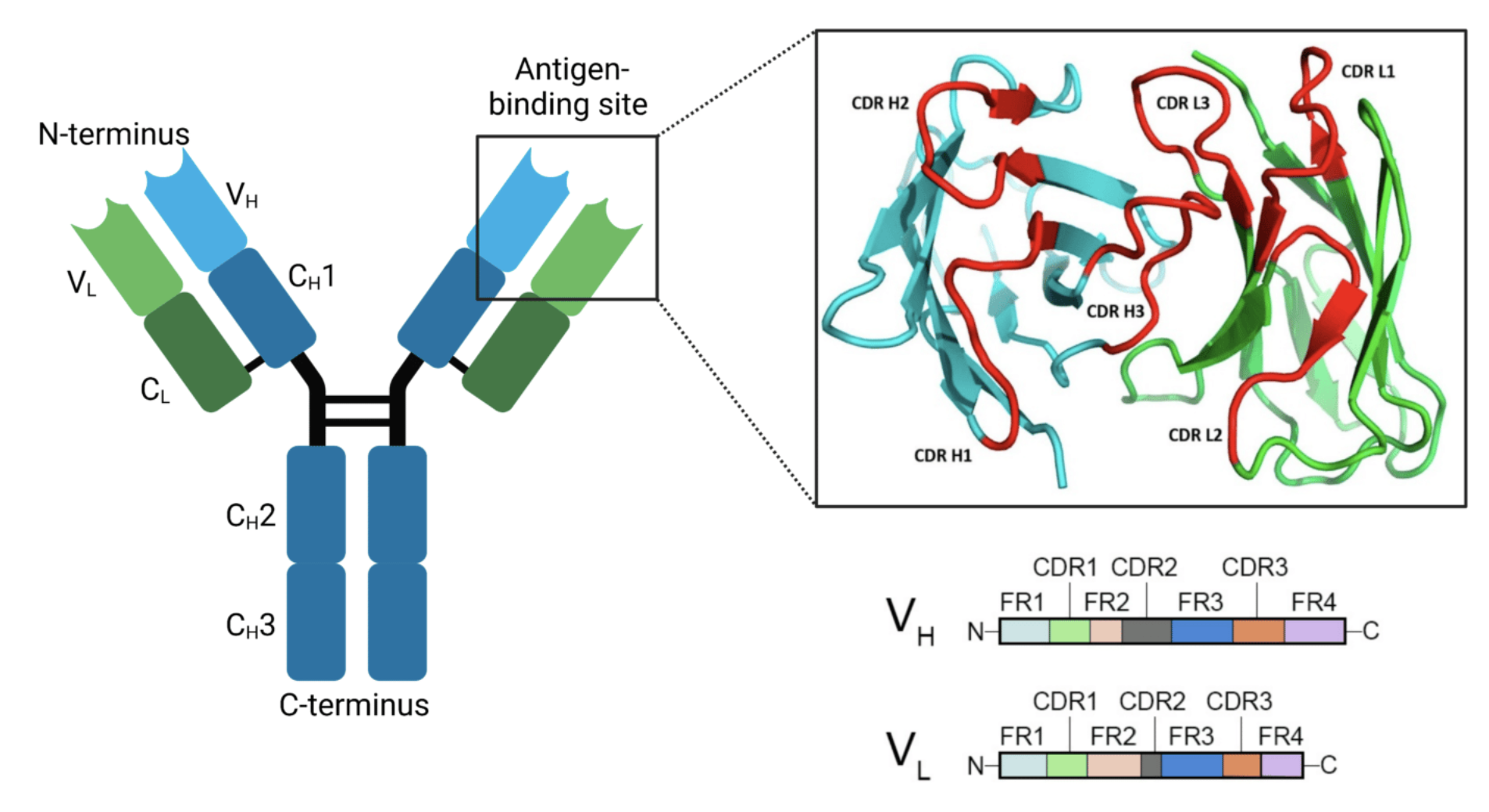}
    \caption{The antigen binding site of an antibody contains CDR and FR regions. CDR regions (L1-3 and H1-3) make up the antigen binding site on the N-terminus of the antibody.}
    \label{fig:ab_struct2}
\end{figure}

\section{Dataset Links and Documentation}

\subsection{Data documentation}
We use data card to document our dataset, as suggested by \citep{pushkarna2022data}. Please find the markdown file 
\href{https://github.com/prescient-design/antibody-domainbed}{here}, and, in the attached supplementary material for this submission.

\subsection{URL for download and DOI}
Our dataset on therapeutic antibody designs can be downloaded from Zenodo. Please use the following \href{https://zenodo.org/uploads/11446107}{link}.

Our DOI identifier is \verb|https://doi.org/10.5281/zenodo.11446107|.

\subsection{URL to Croissant metadata}
Please find our croissant metadata alogn with the code in out repository \href{https://github.com/prescient-design/antibody-domainbed}{here}.

\subsection{Author statement}
We, the authors, hereby declare that we bear full responsibility for any violations of rights associated with this submission, including but not limited to intellectual property rights, privacy rights, and any other applicable laws. We confirm that all data used in this manuscript is appropriately licensed and that we have adhered to all relevant data usage and sharing regulations. We have ensured that our data sources comply with legal and ethical standards, and we take full accountability for any discrepancies or issues that may arise in relation to the data presented in this work.

\subsection{Hosting, licencing and maintenance plan}

Antibody Domainbed is made available under the CC BY 4.0 license. A copy of the license is provided
with the dataset. We, the authors bear all responsibility in case of violation of rights.

For detailed information and maintenance plan, please see our data card \href{https://github.com/prescient-design/antibody-domainbed}{here}, or in the submitted supplementary material.

\newpage

\end{document}